\documentclass[twocolumn,showpacs,preprintnumbers,amsmath,amssymb,prb,superscriptaddress]{revtex4}

\usepackage{mciteplus}
\usepackage{graphicx}

\newcommand{\e}{\text{e}}
\newcommand{\ic}{\text{i}}
\newcommand{\dagind}[2]{#1_{#2}^{\vphantom{\dagger}}}

\begin{document}

\title{Long transient dynamics in the Anderson-Holstein model out of equilibrium}
\author{K. F. Albrecht}
\affiliation{Physikalisches Institut, Albert--Ludwigs--Universit\"at Freiburg, Hermann-Herder-Str.~3, D--79104 Freiburg, Germany}
\affiliation{Freiburg Institute for Advanced Studies, Albert--Ludwigs--Universit\"at Freiburg, D--79104 Freiburg, Germany}
\author{A. Martin-Rodero}
\affiliation{Departamento de F\'{\i}sica Te\'{o}rica de la Materia Condensada and Centro de Investigaci\'{o}n de F\'{\i}sica de la Materia Condensada and Instituto Nicol\'{a}s Cabrera, Universidad Aut\'{o}noma de Madrid, E--28049 Madrid, Spain}
\author{R. C. Monreal}
\affiliation{Departamento de F\'{\i}sica Te\'{o}rica de la Materia Condensada and Centro de Investigaci\'{o}n de F\'{\i}sica de la Materia Condensada and Instituto Nicol\'{a}s Cabrera, Universidad Aut\'{o}noma de Madrid, E--28049 Madrid, Spain}
\author{L. M\"uhlbacher}
\affiliation{Physikalisches Institut, Albert--Ludwigs--Universit\"at Freiburg, Hermann-Herder-Str.~3, D--79104 Freiburg, Germany}
\author{A. Levy Yeyati}
\affiliation{Departamento de F\'{\i}sica Te\'{o}rica de la Materia Condensada and Centro de Investigaci\'{o}n de F\'{\i}sica de la Materia Condensada and Instituto Nicol\'{a}s Cabrera, Universidad Aut\'{o}noma de Madrid, E--28049 Madrid, Spain}
\date{\today}
\begin{abstract}
  We calculate the time dependent nonequilibrium current through a single level quantum dot strongly coupled to a vibrational mode. The nonequilibrium real time dynamics caused by an instantaneous coupling of the leads to the quantum dot is discussed using an approximate method. The approach, which is specially designed for the strong polaronic regime, is based on the so-called polaron tunneling approximation. Considering different initial dot occupations, we show that a common steady state is reached after times much larger than the typical electron tunneling times due to a polaron blocking effect in the dot charge. A direct comparison is made with numerically exact data, showing good agreement for the time scales accessible by the diagrammatic Monte Carlo simulation method. 
\end{abstract}
\maketitle
\section{Introduction} \label{introduction}
Experimental progress in the last few years has enabled a detailed study of transport phenomena in single-molecule junctions. \cite{Park02,Liang02,Gaudioso00,Osorio10,Riel06,Choi06} Such a junction can be considered as a quantum dot contacted to two electrodes via a tunneling coupling. Applying a finite voltage the electrons tunnel through the quantum dot. The charging of the molecule leads to elastic deformations of
its geometry which causes a coupling between electronic and vibrational degrees of freedom. This gives rise to effects like steps in the current-voltage characteristics\cite{Secker11,Cuevas10,Smit02,Natelson04,Ballmann10} and the formation of sidebands in the excitation spectra. \cite{reed,Park02,Zhitenev02,Muehlbacher08}
\newline
In general such a quantum dot setup can be described by the Anderson-Holstein \cite{Holstein1959,Hewson2002} model.
When one is mainly interested in the effects caused by the vibrational mode of the molecule it is customary to consider a linearly coupled local phonon mode and a single level quantum dot with spinless electrons.
\newline
Depending on the temperature $T$, the coupling strength of the electrodes to the dot $\Gamma$, the electron-phonon interaction $\lambda$ as well as the phonon frequency $\omega_0$, different physical regimes can be distinguished. In the classical regime, $T \gg \Gamma$, the problem can be treated with semi-classical approaches using master equations \cite{PhysRevB.68.205324,Mitra2004}. On the other hand, in the quantum regime, $T \ll \Gamma$, a great theoretical effort has been made to develop methods to describe transport phenomena within this model.
This includes different approximate approaches (see for example Refs.~[\onlinecite{PhysRevB.80.041309,carmina2009,carmina2010,alvaro2008,PhysRevB.83.085401,Galperin07,Cuevas10,PhysRevB.80.041307,PhysRevLett.103.136601,0953-8984-23-10-105301,PhysRevB.76.033417}] and references therein) as well as numerically exact methods such as the diagrammatic Monte Carlo simulation (diagMC) \cite{Muehlbacher08,Werner2009,Werner2011,PhysRevB.83.075107}, auxiliary-field quantum Monte Carlo \cite{PhysRevB.72.041301} or the multilayer multiconfiguration time dependent Hartree method \cite{Wang09,Wang11} as well as the iterative path integral summation scheme \cite{PhysRevB.85.121408}. 
\newline
While most of these methods address the steady state behavior of the system, the way how this steady state is built up is not yet well understood. In the strong polaronic regime the transient is only addressed by mean field studies \cite{Riwar2009} or numerical methods. Recent numerical calculations \cite{Ferdinand2012} of the transient dynamics of the current in the Anderson-Holstein model were indicating that in this model there exists a large time scale over which different initial preparations lead to different transport properties which might even lead to a bistable situation. A bistable behavior was predicted previously for the steady state \cite{Gogolin2002,Alexandrov2003,Galperin2005,DAmico08} as well as the time dependent \cite{Riwar2009} regime within a mean field theory. 
\newline
In this paper we address the transient behavior of the Anderson-Holstein model for the strong polaronic regime at $T=0$. Our aim is to develop an approximate method in order to analyze the behavior at very long times which is inaccessible by numerically exact approaches. We are specially interested in understanding how the steady state is built up from different initial conditions for the dot occupation. For this purpose we extend a recently proposed strong coupling approximation, namely the polaron tunneling approximation (PTA) \cite{PhysRevB.83.085401}, to non-stationary situations. The results obtained from this approximation are in very good agreement with the numerical ones obtained by diagMC for the times accessible to the exact method. On the other hand, our approximation shows how the system converges into a steady state solution for much larger time scales, regardless of the initial condition.
\newline
The structure of the paper is the following: In section \ref{model} we briefly introduce the Anderson-Holstein model and then discuss our approach to the time dependent problem in the strongly polaronic regime in section \ref{theory}. The results are discussed in section \ref{results}, where we first address time scales accessible by numerical methods so that a comparison can be made. Finally the long time regime is discussed, and we give a simple interpretation of the polaron blocking mechanism. The paper is closed by some concluding remarks.
\section{The model} \label{model}
The setup consists of two electrodes, left (L) and right (R), which are contacted by a tunneling coupling to an atom or molecule (hereafter called quantum dot) which is modeled by a single electronic level. This level is coupled to a single phonon mode which can be considered as the most relevant vibrational mode of the atom or molecule.
\newline
Such a system can be described by the spinless Anderson-Holstein model given by\cite{Glazman1988,Galperin07}
\begin{align}\label{Eq:Anderson-Holstein_Hamiltonian}
  H
  &=
  H_{\text{D}}
  +
  H_{\text{LR}}
  +
  H_{\text{ph}}
  +
  H_{\text{T}}
  +
  H_{\text{I}}
  \,,
\end{align}
where $H_{\text{D}} = \epsilon_{\text{D}} d^\dagger d$ describes the quantum dot energy level $\epsilon_{\text{D}}$, where $d^{\dagger}$ and $d$ are the electron creation and annihilation operators on the dot. $H_{\text{LR}} = \dagind{\sum}{\alpha, k} \dagind{\epsilon}{\alpha k} a^{\dagger}_{\alpha k} \dagind{a}{\alpha k}$ corresponds to the non-interacting leads, with $\alpha = \text{L,R}$ denoting the left and right electrode respectively. The electronic creation and annihilation operators on electrode $\alpha$ at energy level $\epsilon_{\alpha k}$ are denoted by $a^{\dagger}_{\alpha k}$ and $a_{\alpha k}$. The two electrodes are kept at different chemical potentials via a constant bias voltage $eV=\mu_{\text{L}} - \mu_{\text{R}}$, inducing a nonequilibrium current through the dot. The tunneling coupling to the dot is described by
\begin{align}
  H_{\text{T}}
  &=
  \sum_{\alpha,k} \gamma_{\alpha}
  \left(
    a^{\dagger}_{\alpha k} 
    d 
    + 
    \text{H.c.}
  \right) 
  \,,
\end{align}
where $\gamma_{\alpha}$ are the tunneling amplitudes. We additionally introduce the tunneling rates $\Gamma_{\alpha} = \pi \gamma_{\alpha}^2 \rho_{\alpha}$, where $\rho_{\alpha}$ is the density of states of electrode $\alpha$ which is assumed to be independent of the energy. We further assume $\Gamma_{\text{L}}=\Gamma_{\text{R}}=\Gamma/2$.
\newline
The phonon mode is described by $H_{\text{ph}} = \omega_0 b^{\dagger} b$, which models a vibrational degree of freedom of the molecule with a normal mode of frequency $\omega_0$. The coupling of this mode to the dot level is described by
\begin{align}
  H_{\text{I}}
  &= \lambda
  d^{\dagger} d 
  \left(
    b
    + 
    b^{\dagger}
  \right)
  \,,
\end{align}
with the coupling constant $\lambda$.
Throughout this paper we set $\hbar=e=m_e=1$. 
\section{Green function approach for the polaronic regime}\label{theory}
Despite its simple structure, no exact analytical solution of the Anderson-Holstein model is known for arbitrary parameters. Only in the limits of either a vanishing dot-lead coupling \cite{Mahan1991}, often called the atomic limit, or in the absence of phonons \cite{Schmidt2008} an exact solution can be obtained. For the stationary case an exact solution can also be found in the limits $\epsilon_{\text{D}}/\Gamma \rightarrow \pm \infty$ \cite{hewson-newns}. 
\newline
A common procedure to address the steady state is to assume a decoupled situation in the infinitely past, that is, at $t=-\infty$. Therefore, for any time of interest, the system is in its steady state and transient effects do not need to be treated explicitly, which often simplifies the calculations. Despite the success of this procedure the information about how the steady state is established cannot be gathered. Such transients can be studied assuming an initially vanishing tunneling coupling between the dot and the electrodes. Then the tunneling coupling of the dot to the leads is switched on at a certain initial time $t=0$ and the subsequent time evolution to the steady state can be analyzed. In the present work we develop an analytical approach based on nonequilibrium Green's function techniques to study this transient behavior of the Anderson-Holstein model.
\newline
In order to access the nonequilibrium properties of the dot, the Keldysh Green's functions have to be determined
\begin{align}
  D(t,t^{\prime})
  &=
  -
  \ic
  \left \langle
    \mathcal{T}_{\mathcal{C}}
    d(t)
    d^{\dagger}(t^{\prime})
    X(t)
    X^{\dagger}(t^{\prime})
  \right \rangle
  \label{eq:exact_dots_greens_function}
  \,,
\end{align}
where $\mathcal{T}_{\mathcal{C}}$ denotes the time ordering operator on the Keldysh contour $\mathcal{C}$. The averaging is performed with respect to the complete quantum mechanical state of the system. $X=\e^{\sqrt{g} \left(b^{\dagger} - b \right)}$ is the phonon cloud operator, which is obtained from a unitary Firsov-Lang transformation, \cite{lang_firsov1963} where $g=\left(\lambda/\omega_0\right)^2$. This parameter is a measure of the number of phonons forming the phonon cloud.
\newline
Due to the internal symmetries in Keldysh space \cite{keldysh} it is sufficient to consider only the advanced $D^{\text{a}}(t,t^{\prime})$, retarded $D^{\text{r}}(t,t^{\prime})$, and lesser $D^{<}(t,t^{\prime})$ dot's Green's function to describe the transient current and dot occupation.
\newline
A closed form solution of the complete Green's function in Eq.~(\ref{eq:exact_dots_greens_function}) is hard to achieve since all diagrams containing multi-phonon correlations have to be evaluated explicitly. Therefore, a simple approximation is desirable in order to describe strong electron-phonon couplings in a generic non-stationary regime.
\newline
In this manuscript the time dependence is addressed by a diagrammatic expansion in terms of the dot-lead coupling. The average of the Green's function in Eq.~(\ref{eq:exact_dots_greens_function}) is then defined with respect to some given initial preparation. For the initial preparation two possible dot occupations are considered: Either the dot is empty, so that $n_{\text{D}}(t=0)=0$, or occupied, $n_{\text{D}}(t=0)=1$.
\subsection{Atomic limit}
A good reference for analyzing the strong coupling regime is provided by the atomic limit, defined as the limit in which the tunneling rate $\Gamma$ tends to zero. Following [\onlinecite{Mahan1991}] the model can be solved exactly so that e.~g. the atomic retarded dot's Green's function at zero temperature is given by
\begin{widetext}
\begin{align}
  D_{at}^{\text{r}}(t,t^{\prime})
  &=
  -\ic
  \theta(t-t^{\prime})
  \e^{-g}
  \e^{- \ic \tilde{\epsilon}_{\text{D}} (t-t^{\prime})} 
  \left[
    (1- n_{\text{D}}(t))
    \e^{g \e^{-\ic \omega_0 (t-t^{\prime})}}
    +
    n_{\text{D}}(t)
    \e^{g \e^{\ic \omega_0 (t-t^{\prime})}}
  \right]
  \label{eq:retarded_atomic_greens_function_time_dependent}
  \,,
\end{align}
\end{widetext}
where the polaron shifted energy level of the dot is $\tilde{\epsilon}_{\text{D}}=\epsilon_{\text{D}} - \lambda^2/\omega_0$ and the dot occupation is denoted by $n_{\text{D}}(t) = \langle d^{\dagger}(t) d(t) \rangle$. Notice that for a strictly isolated dot the charge is constant and can only take the values 0 or 1. But when considered as the limiting case of $\Gamma \rightarrow 0$, $n_{\text{D}}(t)$ corresponds to the mean dot occupation of the coupled system \cite{alvaro2008}. This consideration is useful in the following discussion about the self-consistent determination of the dot charge.
\newline
In frequency domain the retarded atomic Green's function has the form
\begin{align}
  \nonumber
  D_{\text{at}}^{\text{r}}(\omega)
  &=
  \e^{-g}
  \sum_{l=0}^{\infty}
  \frac{g^l}{l!}
  \left[
    \frac{
      1
      -
      n_{\text{D}}
    }{
      \omega
      -
      \tilde{\epsilon}_{\text{D}}
      -
      l \omega_0
      +
      i \vartheta
    }
  \right.
  \\
  & \qquad \qquad \qquad
  +
  \left.
    \frac{
      n_{\text{D}}
    }{
      \omega
      -
      \tilde{\epsilon}_{\text{D}}
      +
      l \omega_0
      +
      i \vartheta
    }
  \right]
  \label{eq:retarded_atomic_greens_function_steady_state}
  \,.
\end{align}
where $\vartheta$ is an infinitesimal. 
\subsection{Approximated PTA}
\begin{figure}[h!]
  \begin{center}
    \includegraphics[width=0.475\textwidth]{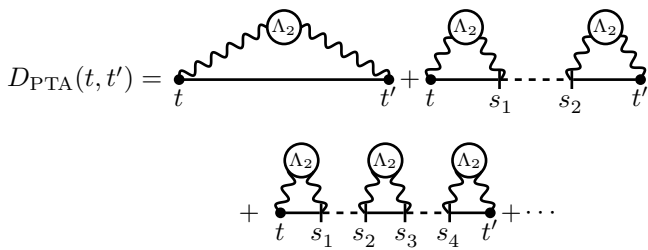}
  \end{center}
  \caption{First three Feynman diagrams of the Dyson series for the PTA. The decoupled dot's Green's function $D_{\text{dot}}(t,t^{\prime})$ is represented with a solid line, the decoupled leads' Green's function $g_{\alpha}(t,t^{\prime})$ with dashed lines. The two point phonon correlator connecting two time integration variables $t,t^ {\prime}$ is denoted by $\Lambda_2 = \left \langle \mathcal{T}_{\mathcal{C}} X(t) X^{\dagger}(t^{\prime}) \right \rangle$.\label{fig:Dyson_series_PTA}}
\end{figure}
\begin{figure}[h!]
  \begin{center}
    \includegraphics[width=0.475\textwidth]{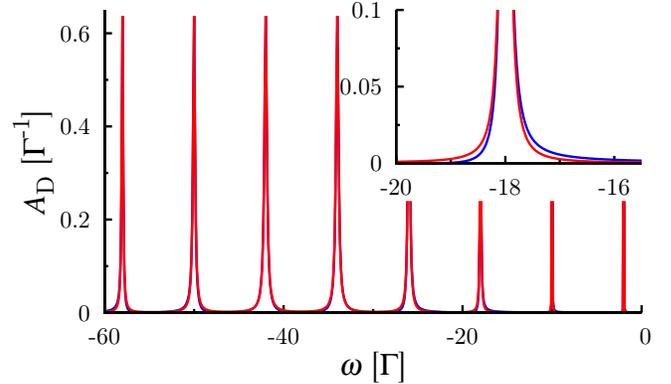}
  \end{center}
  \caption{Spectral density $A_{\text{D}}(\omega)$ of the PTA with fully treating the dot occupation (blue lines) vs. APTA (red lines) for $\tilde{\epsilon}_{\text{D}}=-10 \Gamma$, $\lambda=16\Gamma$, $\omega_0=8\Gamma$ and $V=10\Gamma$. Inset: Zoom of the figure showing the difference between APTA and PTA spectral densities.\label{fig:spect_density}}
\end{figure}

A first step for going beyond the atomic limit is provided by the PTA \cite{PhysRevB.83.085401} which is based on the assumption that the phonons are instantaneously excited once the electron tunnels to the dot and deexcited right after the electron leaves it. Basically, the time scale of an electron on the dot is given by $\tau_{\text{el}} \propto \Gamma^{-1}$. The inverse of the energy due to the polaron formation determines the time scale for an (de)excitation of the polaron, that is, $\tau_{\text{ph}} \propto (\lambda^2/\omega_0)^{-1}$. In the polaronic regime, for $\lambda \gg \Gamma$ and $\lambda/\omega_0 \gg 1$, the lifetime of the electron on the dot is much larger than the (de)excitation time of the polaron $\tau_{\text{el}} \gg \tau_{\text{ph}}$, so that the PTA becomes a reasonable approximation.
\newline
The corresponding Feynman diagrams of the PTA Dyson equation are plotted in Fig.~\ref{fig:Dyson_series_PTA}. Essentially, the PTA replaces multi-polaron correlations by a series of two point correlators $\Lambda_2 = \left \langle \mathcal{T}_{\mathcal{C}} X(t) X^{\dagger}(t^{\prime}) \right \rangle$, that is, all phonon processes with more than one polaron are neglected.
\newline
In the steady state situation the Dyson equation in the Keldysh matrix representation can be solved in frequency space
\begin{align}
  \mathbf{D}_{\text{PTA}} 
  &=
  \mathbf{D}_{\text{at}}
  +
  \mathbf{D}_{\text{at}}
  \mathbf{\Sigma}
  \mathbf{D}_{\text{PTA}}
  \,,
  \label{Dyson-PTA}
\end{align}
by inserting the atomic limit Green's function and the leads' self-energy
\begin{align}
  \mathbf{\Sigma}
  &= 
  \gamma^2 \sigma_z
  \left[
    \mathbf{g_{\text{L}}}
    + 
    \mathbf{g_{\text{R}}}
  \right] \sigma_z
  \label{eq:self_energy_time_steady_state}
  \,,
\end{align}
where $\sigma_z$ is a Pauli matrix in Keldysh space and $\mathbf{g_{\alpha}}$ denotes the Green's functions of the decoupled leads. In this way, e.g. the retarded Green's function can be calculated as
\begin{align}
  D_{\text{PTA}}^{\text{r}}(\omega)
  &= 
  \frac{ D^{\text{r}}_{\text{at}}(\omega)}{ 1 + \ic \Gamma D^{\text{r}}_{\text{at}}(\omega)}
  \label{eq:pta_retarded}
  \,.
\end{align}
From this expression it is straightforward to calculate the self-consistent dot charge using
\begin{align}
  n_{\text{D}}
  &= 
  \frac{1}{2\pi} 
  \sum_{\alpha} 
  \int \text{d} \omega 
  f_{\alpha}(\omega) 
  \mbox{Im} 
  D^{\text{r}}(\omega) \;,
\label{self-consistent-charge}
\end{align}
where $f_{\alpha}(\omega)$ denotes the Fermi distribution on the lead $\alpha$.
The corresponding mean current is then obtained from
\begin{align}
  I
  &= 
  \frac{\pi\Gamma}{2} 
  \int \text{d} \omega 
  \left[ 
    f_{\text{L}}(\omega) 
    - 
    f_{\text{R}}(\omega) 
  \right] 
  \mbox{Im} 
  D^{\text{r}}(\omega) \;.
  \label{eq:steady_state_current}
\end{align}
\begin{figure}[h!]
  \begin{center}
    \includegraphics[width=0.475\textwidth]{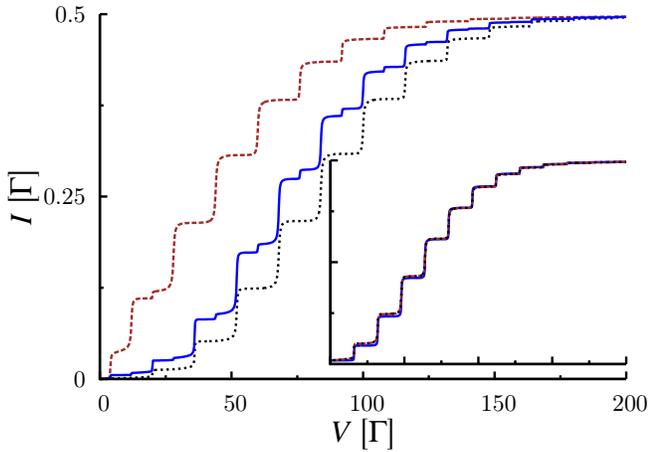}
  \end{center}
  \caption{Steady state current vs. applied bias voltage calculated within the PTA with $n_{\text{D}}=0$ (brown dashed lines), $n_{\text{D}}=1$ (black dotted lines) and with the self-consistent charge (blue lines) determined by Eq.~(\ref{self-consistent-charge}). The dot is off-resonant with $\tilde{\epsilon}_{\text{D}}=-10 \Gamma$, $\lambda=16\Gamma$, $\omega_0=8\Gamma$. Inset: The same plot but for $\tilde{\epsilon}_{\text{D}}= 0$. \label{fig:pta_apta}}
\end{figure}
\\
It is important to notice that the value of the steady-state current in general depends on the value of the mean charge $n_{\text{D}}$. In a non-selfconsistent approach in which the dot charge is assumed to be either 0 or 1 for calculating the retarded Green's function in Eq.~(\ref{eq:pta_retarded}) one would obtain two different values for the stationary current, as illustrated in Fig.~\ref{fig:pta_apta}. However, the self-consistent calculation yields a current-voltage characteristic which lies in between these two results, thus implying the absence of bistability within this approach. A very special situation is that of an electron-hole symmetric case $\tilde{\epsilon}_{\text{D}}=0$, with a symmetric voltage drop. Integrating over this symmetric voltage window in Eq.~(\ref{eq:steady_state_current}) gives an $I-V$ characteristic which does not depend actually on the mean charge (shown as an inset in Fig. \ref{fig:pta_apta}). In the non-symmetric case not only the actual value of the current is different, but also such main features such as the height of the steps at multiples of the phonon frequency and the length of the plateaus between two phonon steps.
\newline
In spite of the simplicity of the expression for the PTA Green's function (Eq.~(\ref{eq:pta_retarded})), its generalization to a time-dependent situation is rather involved. A further simplification of the approximation in the limit $g \gg 1$ and $\omega_0 \gg \Gamma$ can be achieved by noticing that the polaronic (multi-phonon) side-bands essentially do not overlap (see Fig. \ref{fig:spect_density}). This allows to approximate the poles in the retarded PTA Green's function as independent Lorentzian functions
\begin{align}
  \nonumber
  D_{\text{APTA}}^{\text{r}}(\omega) 
  &\approx
  \e^{-g}
  \sum \limits_{l=0}^{\infty}
  \frac{
    g^l
  }{
    l!
  }
  \left(
    \frac{
      1
      -
      n_{\text{D}}
    }{
      \omega
      -
      \tilde{\epsilon}_{\text{D}}
      -
      l \omega_0
      +
      \ic \tilde{\Gamma}_l^-
    }
  \right.
  \\
  & \qquad \qquad \qquad
  \left.
    +
    \frac{
      n_{\text{D}}
    }{
      \omega
      - 
      \tilde{\epsilon}_{\text{D}}
      +
      l\omega_0
      +
      \ic \tilde{\Gamma}_l^+
    }
  \right)
  \,,
\end{align}
where, in order to fit the correct broadening around each resonance, the parameters $\Gamma^{\pm}_l$ have to have the form
\begin{align}
  \tilde{\Gamma}_l^{-}
  &\equiv
  \tilde{\Gamma}_l
  \left[
    \left(
      1
      -
      n_{\text{D}}
    \right)
    +
    n_{\text{D}}
    \delta_{l=0}
  \right]
  \,,
  \\
  \tilde{\Gamma}_l^{+}
  &\equiv
  \tilde{\Gamma}_l
  \left[
    n_{\text{D}}
    +
    (1- n_{\text{D}})
    \delta_{l=0}
  \right]
  \,,
\end{align}
with
\begin{align}
  \tilde{\Gamma}_l
  \equiv
  \Gamma 
  \e^{-g}
  \frac{g^{l}}{l!}
  \,.
\end{align}
This approximated PTA (APTA) gives a much simpler form for the retarded Green's function while all its main features are still preserved. See Fig.~\ref{fig:spect_density} for a comparison of the spectral densities from APTA and PTA in the strong coupling regime.
\\
We further notice that in general the APTA fulfills the PTA Dyson equation (Eq.~(\ref{Dyson-PTA})) approximately, ignoring the cross terms arising from mixing different multiphonon resonances.
\subsection{Time dependent APTA}
In this subsection, we provide an ansatz for describing the time dependent nonequilibrium transient current for two different initial occupations within the spirit of the PTA approach. In principle, one has to solve the time-dependent Dyson equation
\begin{align}
  \nonumber
  \mathbf{D}_{\text{APTA}} (t,t^{\prime}) 
  &=
  \mathbf{D}_{\text{at}} (t,t^{\prime}) 
  + 
  \int \text{d} s_1 
  \int \text{d} s_2 
  \\
  & \quad
  \, 
  \mathbf{D}_{\text{at}} (t,s_1) 
  \,
  \mathbf{\Sigma} (s_1,s_2)  
  \,
  \mathbf{D}_{\text{APTA}} (s_2,t^{\prime}) 
  \label{eq:time_apta_dyson}
  \,,
\end{align}
where the coupling of the dot to the leads is switched on abruptly at $t=0$ via $\gamma(t)=\theta(t) \gamma$. The corresponding self energies then are given by
\begin{align}
  \mathbf{\Sigma} (t,t^{\prime})
  &= 
  \theta(t)
  \theta(t^{\prime})
  \gamma^2 
  \sigma_z \left[
    \mathbf{g}_{\text{L}}(t,t^{\prime})
    + 
    \mathbf{g}_{\text{R}}(t,t^{\prime})
  \right] \sigma_z
  \label{eq:self_energy_time_dependent}
  \,.
\end{align}
A full self-consistent solution of these integral equations is a formidable task. The main idea of our ansatz is to perform a quasi-adiabatic approach in which the charge of the dot is assumed to evolve slowly in time while the spectral density adapts to the instantaneous value of this charge. This is consistent with the general PTA picture, since the average occupation of the quantum dot changes on a time scale given by $\Gamma^{-1}$ whereas the phononic degrees of freedom adapt to the dot occupation on a time scale given by $(\lambda^2/\omega_0)^{-1}$. This permits a reasonable "closed" or compact solution of the dynamical problem. Technically, we shall neglect the transient effects in the retarded Green's functions while focusing on the transient properties of the lesser Green's function, which allows to determine the time-dependent charge self-consistently. More explicitly, for the retarded Green's function our ansatz for $t,t^{\prime}>0$ is
\begin{widetext}
  \begin{align}
    D_{\text{APTA}}^{\text{r}}(t,t^{\prime})
    &=
    -\ic
    \theta (t-t^{\prime})
    \e^{-g}
    \e^{- i \tilde{\epsilon}_{\text{D}} (t-t^{\prime})} 
    \sum \limits_{l=0}^{\infty}
    \frac{g^{l}}{l!}
    \left[
      (1-n_{\text{D}}(t))
      \e^{- \tilde{\Gamma}_l^{-}(t) (t-t^{\prime})}
      \e^{-\ic \omega_0 l (t-t^{\prime})}
      +
      n_{\text{D}}(t)
      \e^{- \tilde{\Gamma}_l^{+}(t) (t-t^{\prime})}
      \e^{\ic \omega_0 l (t-t^{\prime})}
    \right]
    \label{eq:retarded_apta_ansatz}
    \,,
  \end{align}
\end{widetext}
where, similarly to the steady state, the side-band broadenings have the form
\begin{align}
  \tilde{\Gamma}_l^{-}(t)
  &\equiv
  \tilde{\Gamma}_l
  \left[
    \left(
      1
      -
      n_{\text{D}}(t)
    \right)
    +
    n_{\text{D}}(t)
    \delta_{l=0}
  \right]
  \,,
  \\
  \tilde{\Gamma}_l^{+}(t)
  &\equiv
  \tilde{\Gamma}_l
  \left[
    n_{\text{D}}(t)
    +
    (1-n_{\text{D}}(t))
    \delta_{l=0}
  \right]
  \,,
\end{align}
with $n_{\text{D}}(t)$ being the instantaneous mean charge which has to be determined self-consistently from the lesser Green's function
\begin{align}
  n_{\text{D}}(t)
  =
  -\ic
  D_{\text{APTA}}^<(t,t)
  \ .
\end{align}
This Green's function satisfies the corresponding Dyson equation
\begin{align}
  \nonumber
  D^<_{\text{APTA}} 
  &= 
  \left(
    1
    + 
    D_{\text{APTA}}^{\text{r}} 
    \Sigma^{\text{r}}
  \right) 
  D_{\text{0}}^< 
  \left(
    1
    + 
    \Sigma^{\text{a}} 
    D_{\text{APTA}}^{\text{a}}
  \right) 
  \\
  & \qquad
  + 
  D_{\text{APTA}}^{\text{r}} 
  \Sigma^< 
  D_{\text{APTA}}^{\text{a}} 
  \,,
\label{Dyson-lesser}
\end{align}
where integration over internal time arguments is implicitly assumed. The advanced dot's Green's function needed for this Dyson equation can be obtained from the retarded one by the general relation $D^{\text{a}}(t,t^{\prime})=\left( D^{\text{r}}(t^{\prime},t) \right)^{\ast}$. In Eq.~(\ref{Dyson-lesser}) the initial condition is provided by $D^{<}_{\text{0}}$ which is determined by the initial dot occupation 
\begin{align}
  D_{\text{0}}^<(t-t^{\prime})
  &=
  \ic
  \e^{ -g }
  \e^{-\ic \tilde{\epsilon}_{\text{D}} (t-t^{\prime})} 
  n_{\text{D}}(0)
  \e^{ g \e^{\ic \omega_0 (t-t^{\prime})}} 
  \label{eq:lesser_atomic_greens_function_time_dependent}
  \,.
\end{align}
Finally, the self-energies in Eq.~(\ref{Dyson-lesser}) are given by
\begin{align}
  \Sigma^{\text{r}}(t,t') &= -\ic \theta(t) \Gamma \delta(t-t') 
  \,,
  \\
  \Sigma^<(t,t') &= \ic \theta(t)\theta(t') 
  \frac{\Gamma}{2 \pi}
  \sum_{\alpha} \int
  \text{d} \omega
  \e^{-\ic \omega(t-t')} f_{\alpha}(\omega)
  \,.
\end{align}
Next, the two possible initial occupations $n_{\text{D}}(0)=0$ and $n_{\text{D}}(0)=1$ are discussed separately. In the first case only the second term on the rhs of Eq.~(\ref{Dyson-lesser}) contributes and the resulting time dependent dot occupation is
\begin{widetext}
  \begin{align}
    n_{\text{D}}(t)
    &=
    \frac{\Gamma}{2 \pi}
    \sum_{\alpha=\text{L,R}}
    \int \limits_{-\infty}^{\infty}
    \text{d} \omega  f_{\alpha}(\omega)
    \left|
      (1 - n_{\text{D}}(t)) S_{-}(\omega,t)
      + 
      n_{\text{D}}(t)
      S_{+}(\omega,t)
    \right|^2
    \,,
    \label{eq:time_dependent_dot_occupation_apta}
  \end{align}
\end{widetext}
where  
\begin{align}
  S_{\pm}(\omega,t)
  &= \e^{-g} \sum_{l=0}^{\infty} \left(\frac{g^l}{l!}\right) 
  \frac{
    \e^{
      -\ic
      \left(
        \omega
        -
        \tilde{\epsilon}_{\text{D}}
      \right)
      t
    }
    -
    \e^{- \tilde{\Gamma}_l^{\pm}(t) t}
    \e^{\pm \ic \omega_0 lt} 
  }{
    \omega
    -
    \tilde{\epsilon}_{\text{D}}
    \pm
    \omega_0 l
    +
    \ic
    \tilde{\Gamma}_l^{\pm}(t)
  }
  \label{eq:time_dependent_apta_S_function}
  \,.
\end{align}
On the other hand, for the case when $n_{\text{D}}(0)=1$, there is an extra contribution $\delta n_{\text{D}}(t)$ arising from the first term on the rhs of Eq.~(\ref{Dyson-lesser}) given by 
\begin{align}
  \nonumber 
  \delta n_{\text{D}}(t)
  & =
  n_{\text{D}}(0) 
  \e^{-g} 
  \sum_{l=0}^{\infty} 
  \frac{g^l}{l!}
  \\
  & \quad
  \times
  \left| 
    1 
    + 
    \left(
      1
      - 
      n_{\text{D}}(t)
    \right) 
    A^+_l(t)  
    + 
    n_{\text{D}}(t) 
    A^-_l(t) 
  \right|^2 
  \label{eq:time_dependent_dot_occupation_apta_occ}
  \,,
\end{align}
where
\begin{align}
  A^{\pm}_l(t) 
  &= 
  \ic 
  \sum_{m=0}^{\infty} 
  \tilde{\Gamma}_m
  \frac{
    1 
    - 
    \e^{- \tilde{\Gamma}_l^{\pm}(t) t}
    \e^{-\ic \omega_0 (l\mp m) t}
    }{
      (l \mp m) \omega_0 
      - 
      \ic \tilde{\Gamma}_l^{\pm}(t)
    }
  \,.
\end{align}
Reaching a unique stationary state would require that $\delta n_{\text{D}}(t) \rightarrow 0$ for $t \rightarrow \infty$. Although this would be warranted in an exact time-dependent self-consistent PTA, the approximations done within APTA yield a small finite correction which vanishes as $\tilde{\Gamma}/\omega_0$ tends to zero. In order to numerically evaluate the dot occupation, a finite time step is chosen and the equations for the dot occupations (Eqs.~(\ref{eq:time_dependent_dot_occupation_apta}) and (\ref{eq:time_dependent_dot_occupation_apta_occ})) are solved iteratively starting from the initial condition $n_{\text{D}}(t=0) = \{ 0 , 1\}$. 
\newline
The average current is calculated using the relation
\begin{align}
  \nonumber
  I_{\text{av}}(t)
  &=
  \frac{1}{2}
  \left[
    I_{\text{L}}(t)
    -
    I_{\text{R}}(t)
  \right]
  \\
  &=
  \gamma^2
  \text{Re}
  \int \limits_0^{\infty} \text{d}s
  D^{\text{r}}(t,s) 
  \left[
    g_{\text{L}}^{\text{K}}(s,t) 
    -
    g_{\text{R}}^{\text{K}}(s,t)
  \right]
  \,,
  \label{eq:average_current_symmetric_leads}
\end{align}
where $g_{\alpha}^{\text{K}}(t,t^{\prime})$ denotes the Keldysh Green's function of the decoupled lead $\alpha$. Inserting the Green's functions and performing the time integration one obtains
\begin{widetext}
  \begin{align}
    \nonumber
    I_{\text{av}}(t)
    &=
    \theta(t)
    \langle I \rangle
    +
    \theta(t)
    \frac{1}{2\pi}
    \sum \limits_{l=0}^{\infty}
    \tilde{\Gamma}_l
    \int \limits_{-V/2-\tilde{\epsilon}_{\text{D}}}^{V/2-\tilde{\epsilon}_{\text{D}}}
    \text{d} \omega
    \left \lbrace
      (1-n_{\text{D}}(t))
      \frac{
        \e^{-\tilde{\Gamma}_l^{-}(t) t}
        \left(
          \left(
            \omega_0 l
            - 
            \omega
          \right)
          \sin
          \left(
            \left(
              \omega_0 l
              - 
              \omega
            \right)
            t
          \right)
          -
          \tilde{\Gamma}_l^{-}(t)
          \cos
          \left(
            \left(
              \omega_0 l
              - 
              \omega
            \right)
            t
          \right)
        \right)
      }{
        \left(
          \omega_0 l
          - 
          \omega
        \right)^2
        + 
        \left(
          \tilde{\Gamma}_l^{-}(t)
        \right)^2
      }
    \right.
    \\
    &
    \left.
      \qquad \quad
      +
      n_{\text{D}}(t)
      \frac{
        \e^{-\tilde{\Gamma}_l^{+}(t) t}
        \left(
          \left(
            \omega_0 l
            + 
            \omega
          \right)
          \sin
          \left(
            \left(
              \omega_0 l
              + 
              \omega
            \right)
            t
          \right)
          -
          \tilde{\Gamma}_l^{+}(t)
          \cos
          \left(
            \left(
              \omega_0 l
              + 
              \omega
            \right)
            t
          \right)
        \right)
      }{
        \left(
          \omega_0 l
          + 
          \omega
        \right)^2
        + 
        \left(
          \tilde{\Gamma}_l^{+}(t)
        \right)^2
      }
    \right \rbrace
    \label{eq:apta_time_dependent_I_av}
    \,.
  \end{align}
\end{widetext}
Note, that the case of a non-interacting electronic quantum dot can be obtained by setting $g \to 0$, with the same result for the current and the dot occupation as in reference [\onlinecite{Schmidt2008}].
\section{Results}\label{results}
In this section the time dependent APTA results are discussed and compared with the numerically exact data from the diagrammatic Monte Carlo (diagMC) simulation method. This method uses a diagrammatic expansion in the dot-lead tunneling coupling. The occurring time integrals are evaluated stochastically using a Monte Carlo algorithm. Time dependent observables such as current or dot occupation can be calculated for arbitrary system parameters like coupling strength, voltage and temperature. Details of the diagMC method can be found for example in [\onlinecite{Muehlbacher08,Werner2011}].
\newline
Despite being numerically exact the diagMC has a drawback since it suffers from the so-called ``sign problem'': The stochastic Monte Carlo sum has to be performed over terms with alternating signs. This leads to large statistical errors in the observables causing the CPU time to scale exponentially with the system time. Therefore, for any realistic setup it is only possible to simulate the time dependent observables up to system times of the order of $10 \Gamma^{-1}$.
\newline
In order to check the reliability of our approach in subsection \ref{short_times} a comparison is made between the APTA and diagMC for times which are accessible by the latter method. The long time scales are discussed in subsection \ref{long_times}.
\newline
In all cases we show results for both the initially empty and occupied dot.
\subsection{Short time scales}\label{short_times}
\begin{figure}[h!]
  \begin{center}
    \includegraphics[width=.45\textwidth]{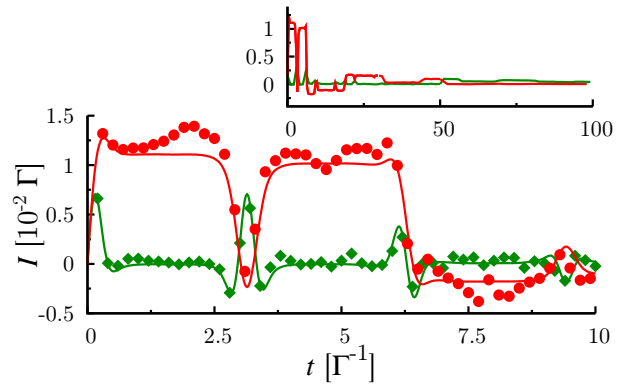}
  \end{center}
  \caption{Comparison between the current from diagMC (symbols) and APTA (straight lines) for $\tilde{\epsilon}_{\text{D}}=-10\Gamma$, $\lambda=16\Gamma$, $\omega_0=8\Gamma$, $V=2\Gamma$. The current from the initially empty (occupied) dot are highlighted in red (green) color for the APTA and represented by dots (diamonds) for the diagMC. Inset: Same plot but for larger times.\label{fig:v1_short_talk_apta}}
\end{figure}
\begin{figure}[h!]
  \begin{center}
    \includegraphics[width=0.475\textwidth]{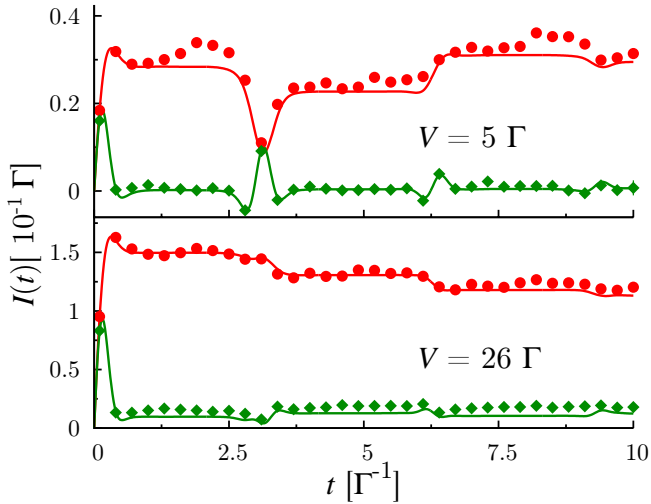}
  \end{center}
  \caption{The currents from diagMC and APTA are compared with the same color code and parameters as in Fig.~\ref{fig:v1_short_talk_apta} but with different voltages $V=5\Gamma$ (figure on the top) and $V=26\Gamma$ (bottom). The steady state current cannot be reached even for times of the order of $10\Gamma^{-1}$.\label{fig:v2.5_short_apta}}
\end{figure}
\begin{figure}[!ht]
  \begin{center}
    \includegraphics[width=0.475\textwidth]{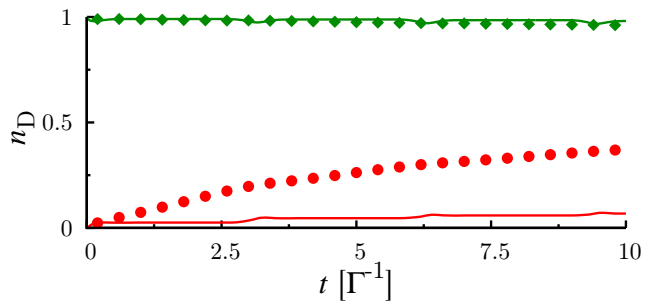}
  \end{center}
  \caption{The dot occupation for the diagMC and APTA are compared for $V=26\Gamma$. While for the dot occupation of the initially occupied dot a good agreement is observed, the APTA dot occupation of the initially empty dot is overestimating the blocking effect with respect to the exact diagMC results.\label{fig:v13_paper_diagMC_apta_short_times_dot_population}}
\end{figure}
\begin{figure}[!h]
  \begin{center}
    \includegraphics[width=0.475\textwidth]{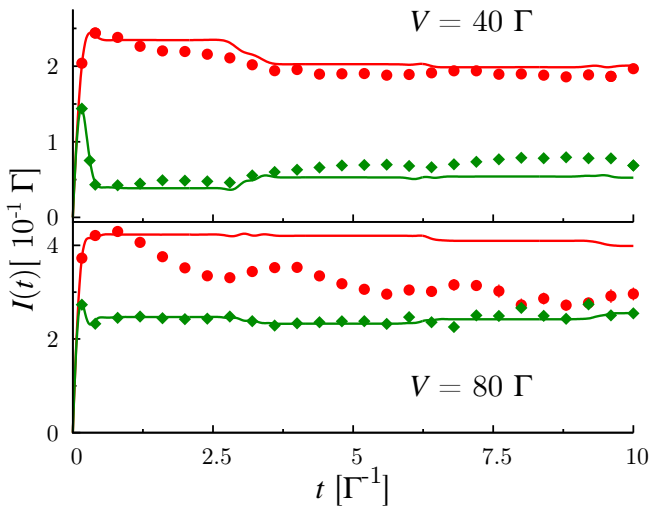}
  \end{center}
  \caption{The current from diagMC and APTA are compared with the same color code and parameters as in Fig.~\ref{fig:v1_short_talk_apta} but with different voltages, $V=40\Gamma$ (top panel) and $V=80\Gamma$ (bottom panel). \label{fig:v20_short_talk_apta}}
\end{figure}

The APTA is expected to be valid in the strong polaronic regime with not too large applied voltages where many-polaron correlations should be small. Therefore, we perform the comparison between APTA and diagMC in the polaronic regime with $\lambda=16\Gamma$, $\omega_0=8\Gamma$ and $\tilde{\epsilon}_{\text{D}}=-10\Gamma$. The choice of these parameters is also guided by the observations of Ref.~[\onlinecite{Ferdinand2012}] suggesting a strong "bistable" like behavior of the system for this case. The voltages are increased from small values to large ones where inelastic processes, not included in the PTA picture, become important.
\newline
The transient currents with $V=2\Gamma$ for the two different initial occupations of the dot, empty or occupied, are plotted in Fig.~\ref{fig:v1_short_talk_apta}, where a remarkable agreement between the APTA and the diagMC is observed. The APTA describes the main transient behavior: The current from APTA is forming plateaus with a constant current which are followed by short time intervals with a rapid change. These two situations exchange each other with a period only depending on the frequency of the phonon. These large oscillations of the current can be interpreted as a shake up process due to the sudden connection of the leads to the dot. For larger times the phonon cloud relaxes and the oscillations become gradually smaller. 
\newline
The transient dynamics in Fig.~\ref{fig:v1_short_talk_apta} are quite different for the two possible initial configurations, empty or occupied. Depending on the initial configuration one observes a peak or a dip at $t=2 \pi n/ \omega_0$, where $n$ is an integer. For times $t \gtrsim 6 \Gamma^{-1}$ both transient currents oscillate around their joint steady state value (see inset of Fig.~\ref{fig:v2.5_short_apta}). Therefore, the time scale for reaching a unique steady state is of the order of several $\Gamma^{-1}$.
\\
Increasing the voltage to $V=5\Gamma$ the time dependent behavior changes: The transient current of the initially empty dot is much larger than the current for the initially occupied one even for the largest times accessed by diagMC (see upper plot in Fig.~\ref{fig:v2.5_short_apta}). Clearly, no joint steady state will be reached within times of the order of $10\Gamma^{-1}$ which is different to the observations for other quantum dot systems such as the Anderson impurity model\cite{Schmidt2008}. Therefore, this effect can be identified as a pure phononic one. The phonons in this regime seem to block the current depending on the initial configuration as it was shown previously in Ref. [\onlinecite{Ferdinand2012}]. 
\newline
The bottom panel of Fig.~\ref{fig:v2.5_short_apta} shows a situation, where the voltage is set to $V=26\Gamma$. Here, the blocking effect is clearly visible since the current is only slightly changing in time but has completely different values depending on the initial preparation even for times of the order of $10 \Gamma^{-1}$. The influence of the phonon shake up process is becoming smaller since more phonon modes contribute due to the increased voltage window.
\newline
The polaron blocking effect observed in the current should also be present in the dot occupancy. In fact, within the APTA the two quantities are intimately connected. In Fig.~\ref{fig:v13_paper_diagMC_apta_short_times_dot_population} the dot population for $V=26\Gamma$ is shown for diagMC and APTA. For the time accessed here, the dot occupation obtained from APTA is only changing slightly so that the initial configuration is preserved even for times of the order of $t > 10 \Gamma^{-1}$.
\newline
The numerical data for the time dependent dot occupation of the initially occupied dot are matching with a high accuracy the APTA results. On the other hand, clear deviations can be seen for the initially empty dot. The APTA seems to overestimate in this case the time scale for the evolution of the charge. 
\\
One reason for this deficiency is that PTA underestimates the width of the resonances in the spectral density far from the Fermi energy \cite{carmina2010}. This deficiency is less important for the evaluation of the current as it is determined by the resonances within the energy window imposed by the electrodes' chemical potentials. An additional source for the discrepancy can be related to the finite bandwidth which was used for the numerical simulation. In contrast to the analytical approach, for a numerical evaluation it is necessary to truncate the density of states in the leads at some finite energy. Since electron transport far away from the Fermi level is important for the time dependent dot occupation it can be strongly influenced by such a finite energy cutoff. This explanation is consistent with the findings of Ref.~[\onlinecite{Schmidt2008}] where a strong dependence of the time dependent dot occupation on the size of the bandwidth was seen in the Anderson impurity model. 
\newline
Further increasing the voltage to $V=40\Gamma$ multi-polaron processes become more important. This leads to small deviation between the APTA results and the diagMC data as it can be seen in Fig.~\ref{fig:v20_short_talk_apta}. Still, the APTA provides a qualitative description showing that the transient currents neither reach a joint steady state, nor a plateau value within the time of several $10\Gamma^{-1}$. 
\\
Finally at $V=80\Gamma$ the deviation between the two calculations becomes more pronounced. While the diagMC results for this bias voltage suggest a convergence towards a steady state on shorter time scales, the APTA for the initially empty dot exhibits a slower convergence. On the other hand, the APTA current for the initially occupied dot still reproduces fairly well the numerically exact result.
\newline
In any case it should be remarked that the APTA predicts in general a slower convergence to the steady state than the diagMC results as it can be seen in Figs.~\ref{fig:v13_paper_diagMC_apta_short_times_dot_population} and \ref{fig:v20_short_talk_apta}. This deficiency, which is more pronounced at larger voltages, can be traced to the already mentioned limitation of the PTA spectral density. 
\subsection{Long time scales}\label{long_times}
For the short time scales analyzed in the previous section it was shown that the phonons induce a blocking effect which leads to a different time-evolution depending on the initial preparation. Further, we have shown that for small to intermediate voltages the APTA describes correctly the transient behavior of the current for the times accessible by the numerically exact diagMC. In this subsection we use APTA to analyze the long time behavior inaccessible to numerically exact simulations.
\begin{figure}[h!]
  \begin{center}
    \includegraphics[width=0.475\textwidth]{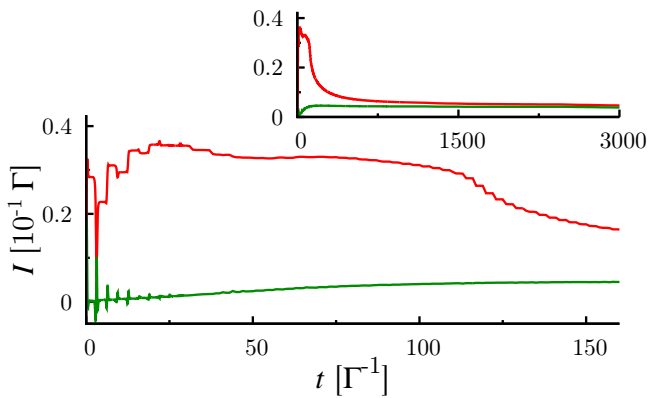}
  \end{center}
  \caption{Time dependent current for long times with the same parameters as in Fig.~\ref{fig:v1_short_talk_apta} and with $V=5\Gamma$. A plateau value of the current for the initially empty dot is observed between $t\approx 20 - 100 \Gamma^{-1}$. Inset: The same plot but for much longer times. \label{fig:v5_scapta}}
\end{figure}
\\
We first focus on the current for the case $V=5\Gamma$ which is plotted in Fig.~\ref{fig:v5_scapta}. Here, the transient currents from different initial preparations are separated at times much larger than several $\Gamma^{-1}$. The polaron blocking effect is clearly visible leading to a slowly varying and almost constant current for each initial occupation up to times $t \approx 100 \Gamma^{-1}$. For larger times, the polaron blocking is no longer the dominant effect, which leads to a charging of the dot and the two currents start to converge. Finally a joint steady state is reached for 
much larger times (See inset in Fig. \ref{fig:v5_scapta}).
\begin{figure}[h!]
  \begin{center}
    \includegraphics[width=0.475\textwidth]{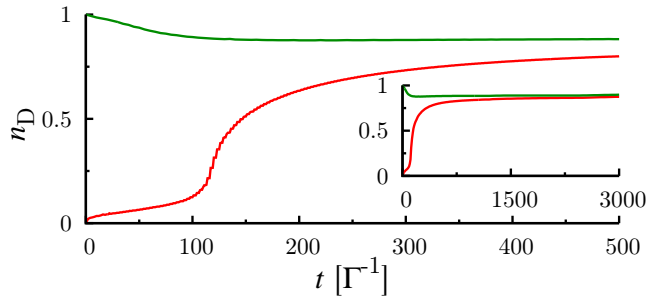}
  \end{center}
  \caption{Time dependent dot occupation for the same parameters as in Fig.~\ref{fig:v5_scapta}. The charge is blocked for the initially empty dot, leading to a small change in the dot occupation up to $t\approx 100\Gamma^{-1}$. Inset: The same plot for a larger time scale. \label{fig:v5_scapta_population}}
\end{figure}
\\
The evolution of the dot occupations for the same values of the parameters as in Fig.~\ref{fig:v5_scapta} is shown in Fig.~\ref{fig:v5_scapta_population}. The charge of the initially occupied dot is close to its steady state value so that the transient behavior is not pronounced. In contrast, for the initially empty dot the occupation is slowly changing until $t \simeq 100 \Gamma^{-1}$. Then a rapid increase is observed and the dot occupation tends towards its stationary state value. 
\newline
This behavior can be understood from Eqs.~(\ref{eq:time_dependent_dot_occupation_apta}) and (\ref{eq:time_dependent_apta_S_function}), giving the evolution of the charge with time. In fact, since $\tilde{\Gamma}_l \ll \Gamma$, $n_{\text{D}}(t)$ can be approximated by
\begin{align}
  n_{\text{D}}(t) 
  &\simeq 
  \frac{1}{2\Gamma} \tilde{\Gamma}^-_1(t) 
  \left(
    1 
    - 
    e^{-\tilde{\Gamma}_1^-(t) t}
  \right)^2 
  \nonumber
  \\
  & \quad + 
  \frac{1}{\Gamma} 
  \sum_{l=0}^{\infty} 
  \tilde{\Gamma}_l^+(t) 
  \left(
    1 
    - 
    e^{-\tilde{\Gamma}_l^+(t) t}
  \right)^2
  \,,
\end{align}
which explicitly exhibits the fact that only one phonon resonance (corresponding to the term in $\tilde{\Gamma}^-_1$) lies within the voltage window in this small voltage range ($V \sim 5\Gamma$). On the other hand, the terms proportional to $\tilde{\Gamma}^+_l(t)$ arise from the occupied resonances below this window. In the case of an initially empty dot, the charge starts to increase with time dominated by the term in $\tilde{\Gamma}_1^{-}(t)$ which behaves as $\tilde{\Gamma}_1^3 (1-n_{\text{D}}(t))^3 t^2$ at short times. The other terms in the above equation give contributions proportional to $n_{\text{D}}(t)^3 t^2$ and are therefore negligible at initial times. When time increases an exponential regime is reached when the terms in $\tilde{\Gamma}_l^{+}(t)$ become important. This change is rather abrupt and happens at times of the order of $\tilde{\Gamma}^+_l(t) t \sim 1$, which roughly corresponds to $t \simeq 100 \Gamma^{-1}$ for the dominant term $l=4$. As the voltage increases and more resonances enter in the voltage window, the charge of the dot increases more quickly and therefore the transition to the exponential behavior occurs at shorter times. This leads to a highly non-linear behavior of the charge both as a function of time and voltage.
\newline
The currents for $V=26\Gamma$ are showing a similar long time behavior (see Fig.~\ref{fig:v13_scapta}). After the current oscillations from the phonon shake up process die out, the current remains almost constant, it only changes slightly in time until $t \approx 50\Gamma^{-1}$ is reached. For larger times a relatively rapid charging of the dot causes the currents to finally reach their joint steady state in a similar way as in the case with $V=5\Gamma$.
\begin{figure}[h!]
  \begin{center}
    \includegraphics[width=0.475\textwidth]{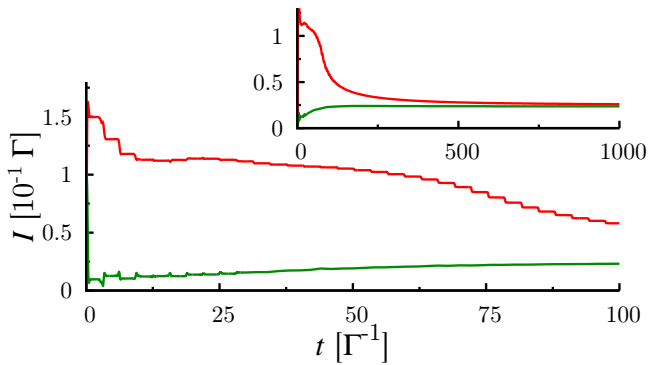}
  \end{center}
  \caption{Time dependent current with the same parameters as in Fig.~\ref{fig:v5_scapta} but with $V=26\Gamma$. The length of the plateau is getting smaller, but still it is clearly visible between $t\approx 10 - 50 \Gamma^{-1}$. Inset: The same plot for larger times.\label{fig:v13_scapta}}
\end{figure}
\\
Increasing the voltage to $V = 40\Gamma$ the convergence to the steady state becomes rather monotonous. Still, the time scales involved are much larger than expected for a pure electronic system  [\onlinecite{Schmidt2008}]. For even larger voltages this time scale is further reduced, however, as commented in the previous section, the PTA would not be able to describe
this large bias regime accurately. 
\begin{figure}[h!]
  \begin{center}
    \includegraphics[width=0.475\textwidth]{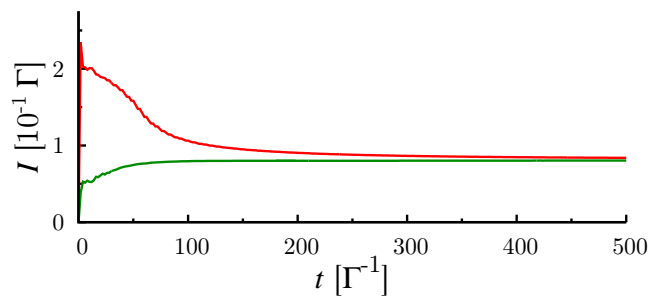}
  \end{center}
  \caption{Time dependent current with the same parameters as in Fig.~\ref{fig:v13_scapta} but with $V=40\Gamma$. The current does not show a plateau for small times but the times until a joint steady state is reached is still large.\label{fig:v20_scapta}}
\end{figure}
\\
Finally, we like to make contact between our theoretical findings and experiments of single molecular junctions by providing a coarse estimate of the set of parameters were we expect that such long transients can be found. Typical values of $\Gamma$ vary between a few $\mu$eVs and several meVs. As an example if we set $\Gamma=$ meV, the parameters would have the values $\lambda=16$ meV, $\omega_0=8$ meV and $\tilde{\epsilon}_{\text{D}}=-10$ meV. The applied bias voltages for which the long time scales are found would then vary between $V=5$ meV and $V=40$ meV. Accordingly, we would obtain transient times of around 50 picoseconds.
\section{Concluding remarks}\label{conclusion}
We have demonstrated the accuracy of the time dependent APTA method in the strong polaronic regime by means of a comparison with the numerically exact diagMC method. A blocking of the current depending on the initial occupation for times of the order of several $\Gamma^{-1}$ was found in agreement to the results of Ref.~[\onlinecite{Ferdinand2012}]. Furthermore, a remarkable agreement with these time dependent results was found up to moderate voltages.  
\newline 
We also used this method to explore the long time scales which are inaccessible for the exact numerical calculations. The polaron blocking effect 
was shown to be connected to the narrowing of the side bands in the spectral density, determined by $\tilde{\Gamma}_l=\Gamma \e^{-g} \frac{g^l}{l!}$ instead of the unrenormalized width $\Gamma$. In this way the time scales of the system can increase by more than one order of magnitude in the polaronic regime. Increasing the voltage, additional side bands contribute to the electronic transport which leads to a faster convergence to the steady state.
\newline
Finally, some limitations of the method developed in this work should be mentioned. Already in the equilibrium case the PTA spectral density underestimates the width of the side bands far from the Fermi level. In a similar way, when a large bias is applied inelastic processes not included in the approximation would become important leading to a faster convergence to the steady state. Further work along this line would be desirable. 
\section*{Acknowledgments}
The authors like to thank A. Komnik and S. Maier for numerous stimulating discussions and H. Wang for providing numerical data. Further the computational resources from the bwGRiD project and financial support by Spanish Mineco through grants FIS2008-04209 and FIS2011-26516 and the European Research Area (ERA) NanoSci Collaborative Project CHENANOM is acknowledged.
\bibliographystyle{apsrevM}
\bibliography{papers}

\ifx\mcitethebibliography\mciteundefinedmacro
\PackageError{apsrevM.bst}{mciteplus.sty has not been loaded}
{This bibstyle requires the use of the mciteplus package.}\fi
\begin{mcitethebibliography}{47}
\expandafter\ifx\csname natexlab\endcsname\relax\def\natexlab#1{#1}\fi
\expandafter\ifx\csname bibnamefont\endcsname\relax
  \def\bibnamefont#1{#1}\fi
\expandafter\ifx\csname bibfnamefont\endcsname\relax
  \def\bibfnamefont#1{#1}\fi
\expandafter\ifx\csname citenamefont\endcsname\relax
  \def\citenamefont#1{#1}\fi
\expandafter\ifx\csname url\endcsname\relax
  \def\url#1{\texttt{#1}}\fi
\expandafter\ifx\csname urlprefix\endcsname\relax\def\urlprefix{URL }\fi
\providecommand{\bibinfo}[2]{#2}
\providecommand{\eprint}[2][]{\url{#2}}

\bibitem[{\citenamefont{Park et~al.}(2002)\citenamefont{Park, Pasupathy,
  Goldsmith, Chang, Yaish, Petta, Rinkoski, Sethna, Abruna, McEuen
  et~al.}}]{Park02}
\bibinfo{author}{\bibfnamefont{J.}~\bibnamefont{Park}},
  \bibinfo{author}{\bibfnamefont{A.}~\bibnamefont{Pasupathy}},
  \bibinfo{author}{\bibfnamefont{J.}~\bibnamefont{Goldsmith}},
  \bibinfo{author}{\bibfnamefont{C.}~\bibnamefont{Chang}},
  \bibinfo{author}{\bibfnamefont{Y.}~\bibnamefont{Yaish}},
  \bibinfo{author}{\bibfnamefont{J.}~\bibnamefont{Petta}},
  \bibinfo{author}{\bibfnamefont{M.}~\bibnamefont{Rinkoski}},
  \bibinfo{author}{\bibfnamefont{J.}~\bibnamefont{Sethna}},
  \bibinfo{author}{\bibfnamefont{H.}~\bibnamefont{Abruna}},
  \bibinfo{author}{\bibfnamefont{P.}~\bibnamefont{McEuen}},
  \bibnamefont{et~al.}, \bibinfo{journal}{Nature (London)}
  \textbf{\bibinfo{volume}{417}}, \bibinfo{pages}{722}
  (\bibinfo{year}{2002})\relax
\mciteBstWouldAddEndPuncttrue
\mciteSetBstMidEndSepPunct{\mcitedefaultmidpunct}
{\mcitedefaultendpunct}{\mcitedefaultseppunct}\relax
\EndOfBibitem
\bibitem[{\citenamefont{Liang et~al.}(2002)\citenamefont{Liang, Shores,
  Bockrath, Long, and Park}}]{Liang02}
\bibinfo{author}{\bibfnamefont{W.}~\bibnamefont{Liang}},
  \bibinfo{author}{\bibfnamefont{M.}~\bibnamefont{Shores}},
  \bibinfo{author}{\bibfnamefont{M.}~\bibnamefont{Bockrath}},
  \bibinfo{author}{\bibfnamefont{J.}~\bibnamefont{Long}}, \bibnamefont{and}
  \bibinfo{author}{\bibfnamefont{H.}~\bibnamefont{Park}},
  \bibinfo{journal}{Nature (London)} \textbf{\bibinfo{volume}{417}},
  \bibinfo{pages}{725} (\bibinfo{year}{2002})\relax
\mciteBstWouldAddEndPuncttrue
\mciteSetBstMidEndSepPunct{\mcitedefaultmidpunct}
{\mcitedefaultendpunct}{\mcitedefaultseppunct}\relax
\EndOfBibitem
\bibitem[{\citenamefont{Gaudioso et~al.}(2000)\citenamefont{Gaudioso, Lauhon,
  and Ho}}]{Gaudioso00}
\bibinfo{author}{\bibfnamefont{J.}~\bibnamefont{Gaudioso}},
  \bibinfo{author}{\bibfnamefont{L.~J.} \bibnamefont{Lauhon}},
  \bibnamefont{and} \bibinfo{author}{\bibfnamefont{W.}~\bibnamefont{Ho}},
  \bibinfo{journal}{Phys. Rev. Lett.} \textbf{\bibinfo{volume}{85}},
  \bibinfo{pages}{1918} (\bibinfo{year}{2000})\relax
\mciteBstWouldAddEndPuncttrue
\mciteSetBstMidEndSepPunct{\mcitedefaultmidpunct}
{\mcitedefaultendpunct}{\mcitedefaultseppunct}\relax
\EndOfBibitem
\bibitem[{\citenamefont{Osorio et~al.}(2010)\citenamefont{Osorio, Ruben,
  Seldenthuis, Lehn, and van~der Zant}}]{Osorio10}
\bibinfo{author}{\bibfnamefont{E.~A.} \bibnamefont{Osorio}},
  \bibinfo{author}{\bibfnamefont{M.}~\bibnamefont{Ruben}},
  \bibinfo{author}{\bibfnamefont{J.~S.} \bibnamefont{Seldenthuis}},
  \bibinfo{author}{\bibfnamefont{J.~M.} \bibnamefont{Lehn}}, \bibnamefont{and}
  \bibinfo{author}{\bibfnamefont{H.~S.~J.} \bibnamefont{van~der Zant}},
  \bibinfo{journal}{Small} \textbf{\bibinfo{volume}{6}}, \bibinfo{pages}{174}
  (\bibinfo{year}{2010})\relax
\mciteBstWouldAddEndPuncttrue
\mciteSetBstMidEndSepPunct{\mcitedefaultmidpunct}
{\mcitedefaultendpunct}{\mcitedefaultseppunct}\relax
\EndOfBibitem
\bibitem[{\citenamefont{L\"ortscher et~al.}(2006)\citenamefont{L\"ortscher,
  Ciszek, Tour, and Riel}}]{Riel06}
\bibinfo{author}{\bibfnamefont{E.}~\bibnamefont{L\"ortscher}},
  \bibinfo{author}{\bibfnamefont{J.~W.} \bibnamefont{Ciszek}},
  \bibinfo{author}{\bibfnamefont{J.}~\bibnamefont{Tour}}, \bibnamefont{and}
  \bibinfo{author}{\bibfnamefont{H.}~\bibnamefont{Riel}},
  \bibinfo{journal}{Small} \textbf{\bibinfo{volume}{2}}, \bibinfo{pages}{973}
  (\bibinfo{year}{2006})\relax
\mciteBstWouldAddEndPuncttrue
\mciteSetBstMidEndSepPunct{\mcitedefaultmidpunct}
{\mcitedefaultendpunct}{\mcitedefaultseppunct}\relax
\EndOfBibitem
\bibitem[{\citenamefont{Choi et~al.}(2006)\citenamefont{Choi, Kahng, Kim, Kim,
  Kim, Song, Ihm, and Kuk}}]{Choi06}
\bibinfo{author}{\bibfnamefont{B.~Y.} \bibnamefont{Choi}},
  \bibinfo{author}{\bibfnamefont{S.~J.} \bibnamefont{Kahng}},
  \bibinfo{author}{\bibfnamefont{S.}~\bibnamefont{Kim}},
  \bibinfo{author}{\bibfnamefont{H.}~\bibnamefont{Kim}},
  \bibinfo{author}{\bibfnamefont{H.~W.} \bibnamefont{Kim}},
  \bibinfo{author}{\bibfnamefont{Y.~J.} \bibnamefont{Song}},
  \bibinfo{author}{\bibfnamefont{J.}~\bibnamefont{Ihm}}, \bibnamefont{and}
  \bibinfo{author}{\bibfnamefont{Y.}~\bibnamefont{Kuk}},
  \bibinfo{journal}{Phys. Rev. Lett.} \textbf{\bibinfo{volume}{96}},
  \bibinfo{pages}{156106} (\bibinfo{year}{2006})\relax
\mciteBstWouldAddEndPuncttrue
\mciteSetBstMidEndSepPunct{\mcitedefaultmidpunct}
{\mcitedefaultendpunct}{\mcitedefaultseppunct}\relax
\EndOfBibitem
\bibitem[{\citenamefont{Secker et~al.}(2011)\citenamefont{Secker, Wagner,
  Ballmann, {H\"artle}, Thoss, and Weber}}]{Secker11}
\bibinfo{author}{\bibfnamefont{D.}~\bibnamefont{Secker}},
  \bibinfo{author}{\bibfnamefont{S.}~\bibnamefont{Wagner}},
  \bibinfo{author}{\bibfnamefont{S.}~\bibnamefont{Ballmann}},
  \bibinfo{author}{\bibfnamefont{R.}~\bibnamefont{{H\"artle}}},
  \bibinfo{author}{\bibfnamefont{M.}~\bibnamefont{Thoss}}, \bibnamefont{and}
  \bibinfo{author}{\bibfnamefont{H.~B.} \bibnamefont{Weber}},
  \bibinfo{journal}{Phys. Rev. Lett.} \textbf{\bibinfo{volume}{106}},
  \bibinfo{pages}{136807} (\bibinfo{year}{2011})\relax
\mciteBstWouldAddEndPuncttrue
\mciteSetBstMidEndSepPunct{\mcitedefaultmidpunct}
{\mcitedefaultendpunct}{\mcitedefaultseppunct}\relax
\EndOfBibitem
\bibitem[{\citenamefont{Cuevas and Scheer}(2010)}]{Cuevas10}
\bibinfo{author}{\bibfnamefont{J.}~\bibnamefont{Cuevas}} \bibnamefont{and}
  \bibinfo{author}{\bibfnamefont{E.}~\bibnamefont{Scheer}},
  \emph{\bibinfo{title}{Molecular Electronics: An Introduction to Theory and
  Experiment}} (\bibinfo{publisher}{World Scientific},
  \bibinfo{address}{Singapore}, \bibinfo{year}{2010})\relax
\mciteBstWouldAddEndPuncttrue
\mciteSetBstMidEndSepPunct{\mcitedefaultmidpunct}
{\mcitedefaultendpunct}{\mcitedefaultseppunct}\relax
\EndOfBibitem
\bibitem[{\citenamefont{Smit et~al.}(2002)\citenamefont{Smit, Noat, Untiedt,
  Lang, van Hemert, and van Ruitenbeek}}]{Smit02}
\bibinfo{author}{\bibfnamefont{R.}~\bibnamefont{Smit}},
  \bibinfo{author}{\bibfnamefont{Y.}~\bibnamefont{Noat}},
  \bibinfo{author}{\bibfnamefont{C.}~\bibnamefont{Untiedt}},
  \bibinfo{author}{\bibfnamefont{N.}~\bibnamefont{Lang}},
  \bibinfo{author}{\bibfnamefont{M.}~\bibnamefont{van Hemert}},
  \bibnamefont{and} \bibinfo{author}{\bibfnamefont{J.}~\bibnamefont{van
  Ruitenbeek}}, \bibinfo{journal}{Nature (London)}
  \textbf{\bibinfo{volume}{419}}, \bibinfo{pages}{906}
  (\bibinfo{year}{2002})\relax
\mciteBstWouldAddEndPuncttrue
\mciteSetBstMidEndSepPunct{\mcitedefaultmidpunct}
{\mcitedefaultendpunct}{\mcitedefaultseppunct}\relax
\EndOfBibitem
\bibitem[{\citenamefont{Yu et~al.}(2004)\citenamefont{Yu, Keane, Ciszek, Cheng,
  Stewart, Tour, and Natelson}}]{Natelson04}
\bibinfo{author}{\bibfnamefont{L.~H.} \bibnamefont{Yu}},
  \bibinfo{author}{\bibfnamefont{Z.~K.} \bibnamefont{Keane}},
  \bibinfo{author}{\bibfnamefont{J.~W.} \bibnamefont{Ciszek}},
  \bibinfo{author}{\bibfnamefont{L.}~\bibnamefont{Cheng}},
  \bibinfo{author}{\bibfnamefont{M.~P.} \bibnamefont{Stewart}},
  \bibinfo{author}{\bibfnamefont{J.~M.} \bibnamefont{Tour}}, \bibnamefont{and}
  \bibinfo{author}{\bibfnamefont{D.}~\bibnamefont{Natelson}},
  \bibinfo{journal}{Phys. Rev. Lett.} \textbf{\bibinfo{volume}{93}},
  \bibinfo{pages}{266802} (\bibinfo{year}{2004})\relax
\mciteBstWouldAddEndPuncttrue
\mciteSetBstMidEndSepPunct{\mcitedefaultmidpunct}
{\mcitedefaultendpunct}{\mcitedefaultseppunct}\relax
\EndOfBibitem
\bibitem[{\citenamefont{Ballmann et~al.}(2010)\citenamefont{Ballmann,
  Hieringer, Secker, Zheng, Gladysz, G\"orling, and Weber}}]{Ballmann10}
\bibinfo{author}{\bibfnamefont{S.}~\bibnamefont{Ballmann}},
  \bibinfo{author}{\bibfnamefont{W.}~\bibnamefont{Hieringer}},
  \bibinfo{author}{\bibfnamefont{D.}~\bibnamefont{Secker}},
  \bibinfo{author}{\bibfnamefont{Q.}~\bibnamefont{Zheng}},
  \bibinfo{author}{\bibfnamefont{J.~A.} \bibnamefont{Gladysz}},
  \bibinfo{author}{\bibfnamefont{A.}~\bibnamefont{G\"orling}},
  \bibnamefont{and} \bibinfo{author}{\bibfnamefont{H.~B.} \bibnamefont{Weber}},
  \bibinfo{journal}{Chem. Phys. Chem.} \textbf{\bibinfo{volume}{11}},
  \bibinfo{pages}{2256} (\bibinfo{year}{2010})\relax
\mciteBstWouldAddEndPuncttrue
\mciteSetBstMidEndSepPunct{\mcitedefaultmidpunct}
{\mcitedefaultendpunct}{\mcitedefaultseppunct}\relax
\EndOfBibitem
\bibitem[{\citenamefont{Reed et~al.}(1997)\citenamefont{Reed, Zhou, Muller,
  Burgin, and Tour}}]{reed}
\bibinfo{author}{\bibfnamefont{M.~A.} \bibnamefont{Reed}},
  \bibinfo{author}{\bibfnamefont{C.}~\bibnamefont{Zhou}},
  \bibinfo{author}{\bibfnamefont{C.~J.} \bibnamefont{Muller}},
  \bibinfo{author}{\bibfnamefont{T.~P.} \bibnamefont{Burgin}},
  \bibnamefont{and} \bibinfo{author}{\bibfnamefont{J.~M.} \bibnamefont{Tour}},
  \bibinfo{journal}{Science} \textbf{\bibinfo{volume}{278}},
  \bibinfo{pages}{252} (\bibinfo{year}{1997})\relax
\mciteBstWouldAddEndPuncttrue
\mciteSetBstMidEndSepPunct{\mcitedefaultmidpunct}
{\mcitedefaultendpunct}{\mcitedefaultseppunct}\relax
\EndOfBibitem
\bibitem[{\citenamefont{Zhitenev et~al.}(2002)\citenamefont{Zhitenev, Meng, and
  Bao}}]{Zhitenev02}
\bibinfo{author}{\bibfnamefont{N.~B.} \bibnamefont{Zhitenev}},
  \bibinfo{author}{\bibfnamefont{H.}~\bibnamefont{Meng}}, \bibnamefont{and}
  \bibinfo{author}{\bibfnamefont{Z.}~\bibnamefont{Bao}},
  \bibinfo{journal}{Phys. Rev. Lett.} \textbf{\bibinfo{volume}{88}},
  \bibinfo{pages}{226801} (\bibinfo{year}{2002})\relax
\mciteBstWouldAddEndPuncttrue
\mciteSetBstMidEndSepPunct{\mcitedefaultmidpunct}
{\mcitedefaultendpunct}{\mcitedefaultseppunct}\relax
\EndOfBibitem
\bibitem[{\citenamefont{M\"uhlbacher and Rabani}(2008)}]{Muehlbacher08}
\bibinfo{author}{\bibfnamefont{L.}~\bibnamefont{M\"uhlbacher}}
  \bibnamefont{and} \bibinfo{author}{\bibfnamefont{E.}~\bibnamefont{Rabani}},
  \bibinfo{journal}{Phys. Rev. Lett.} \textbf{\bibinfo{volume}{100}},
  \bibinfo{pages}{176403} (\bibinfo{year}{2008}),
  \urlprefix\url{http://link.aps.org/doi/10.1103/PhysRevLett.100.176403}\relax
\mciteBstWouldAddEndPuncttrue
\mciteSetBstMidEndSepPunct{\mcitedefaultmidpunct}
{\mcitedefaultendpunct}{\mcitedefaultseppunct}\relax
\EndOfBibitem
\bibitem[{\citenamefont{Holstein}(1959)}]{Holstein1959}
\bibinfo{author}{\bibfnamefont{T.}~\bibnamefont{Holstein}},
  \bibinfo{journal}{Annals of Physics} \textbf{\bibinfo{volume}{8}},
  \bibinfo{pages}{343 } (\bibinfo{year}{1959}), ISSN \bibinfo{issn}{0003-4916},
  \urlprefix\url{http://www.sciencedirect.com/science/article/pii/000349165990%
003X}\relax
\mciteBstWouldAddEndPuncttrue
\mciteSetBstMidEndSepPunct{\mcitedefaultmidpunct}
{\mcitedefaultendpunct}{\mcitedefaultseppunct}\relax
\EndOfBibitem
\bibitem[{\citenamefont{Hewson and Meyer}(2002)}]{Hewson2002}
\bibinfo{author}{\bibfnamefont{A.~C.} \bibnamefont{Hewson}} \bibnamefont{and}
  \bibinfo{author}{\bibfnamefont{D.}~\bibnamefont{Meyer}},
  \bibinfo{journal}{Journal of Physics: Condensed Matter}
  \textbf{\bibinfo{volume}{14}}, \bibinfo{pages}{427} (\bibinfo{year}{2002}),
  \urlprefix\url{http://stacks.iop.org/0953-8984/14/i=3/a=312}\relax
\mciteBstWouldAddEndPuncttrue
\mciteSetBstMidEndSepPunct{\mcitedefaultmidpunct}
{\mcitedefaultendpunct}{\mcitedefaultseppunct}\relax
\EndOfBibitem
\bibitem[{\citenamefont{Braig and Flensberg}(2003)}]{PhysRevB.68.205324}
\bibinfo{author}{\bibfnamefont{S.}~\bibnamefont{Braig}} \bibnamefont{and}
  \bibinfo{author}{\bibfnamefont{K.}~\bibnamefont{Flensberg}},
  \bibinfo{journal}{Phys. Rev. B} \textbf{\bibinfo{volume}{68}},
  \bibinfo{pages}{205324} (\bibinfo{year}{2003}),
  \urlprefix\url{http://link.aps.org/doi/10.1103/PhysRevB.68.205324}\relax
\mciteBstWouldAddEndPuncttrue
\mciteSetBstMidEndSepPunct{\mcitedefaultmidpunct}
{\mcitedefaultendpunct}{\mcitedefaultseppunct}\relax
\EndOfBibitem
\bibitem[{\citenamefont{Mitra et~al.}(2004)\citenamefont{Mitra, Aleiner, and
  Millis}}]{Mitra2004}
\bibinfo{author}{\bibfnamefont{A.}~\bibnamefont{Mitra}},
  \bibinfo{author}{\bibfnamefont{I.}~\bibnamefont{Aleiner}}, \bibnamefont{and}
  \bibinfo{author}{\bibfnamefont{A.~J.} \bibnamefont{Millis}},
  \bibinfo{journal}{Phys. Rev. B} \textbf{\bibinfo{volume}{69}},
  \bibinfo{pages}{245302} (\bibinfo{year}{2004})\relax
\mciteBstWouldAddEndPuncttrue
\mciteSetBstMidEndSepPunct{\mcitedefaultmidpunct}
{\mcitedefaultendpunct}{\mcitedefaultseppunct}\relax
\EndOfBibitem
\bibitem[{\citenamefont{Avriller and Levy~Yeyati}(2009)}]{PhysRevB.80.041309}
\bibinfo{author}{\bibfnamefont{R.}~\bibnamefont{Avriller}} \bibnamefont{and}
  \bibinfo{author}{\bibfnamefont{A.}~\bibnamefont{Levy~Yeyati}},
  \bibinfo{journal}{Phys. Rev. B} \textbf{\bibinfo{volume}{80}},
  \bibinfo{pages}{041309} (\bibinfo{year}{2009})\relax
\mciteBstWouldAddEndPuncttrue
\mciteSetBstMidEndSepPunct{\mcitedefaultmidpunct}
{\mcitedefaultendpunct}{\mcitedefaultseppunct}\relax
\EndOfBibitem
\bibitem[{\citenamefont{Monreal and Martin-Rodero}(2009)}]{carmina2009}
\bibinfo{author}{\bibfnamefont{R.~C.} \bibnamefont{Monreal}} \bibnamefont{and}
  \bibinfo{author}{\bibfnamefont{A.}~\bibnamefont{Martin-Rodero}},
  \bibinfo{journal}{Phys. Rev. B} \textbf{\bibinfo{volume}{79}},
  \bibinfo{pages}{115140} (\bibinfo{year}{2009}),
  \urlprefix\url{http://link.aps.org/doi/10.1103/PhysRevB.79.115140}\relax
\mciteBstWouldAddEndPuncttrue
\mciteSetBstMidEndSepPunct{\mcitedefaultmidpunct}
{\mcitedefaultendpunct}{\mcitedefaultseppunct}\relax
\EndOfBibitem
\bibitem[{\citenamefont{Monreal et~al.}(2010)\citenamefont{Monreal, Flores, and
  Martin-Rodero}}]{carmina2010}
\bibinfo{author}{\bibfnamefont{R.~C.} \bibnamefont{Monreal}},
  \bibinfo{author}{\bibfnamefont{F.}~\bibnamefont{Flores}}, \bibnamefont{and}
  \bibinfo{author}{\bibfnamefont{A.}~\bibnamefont{Martin-Rodero}},
  \bibinfo{journal}{Phys. Rev. B} \textbf{\bibinfo{volume}{82}},
  \bibinfo{pages}{235412} (\bibinfo{year}{2010}),
  \urlprefix\url{http://link.aps.org/doi/10.1103/PhysRevB.82.235412}\relax
\mciteBstWouldAddEndPuncttrue
\mciteSetBstMidEndSepPunct{\mcitedefaultmidpunct}
{\mcitedefaultendpunct}{\mcitedefaultseppunct}\relax
\EndOfBibitem
\bibitem[{\citenamefont{Martin-Rodero et~al.}(2008)\citenamefont{Martin-Rodero,
  Levy~Yeyati, Flores, and Monreal}}]{alvaro2008}
\bibinfo{author}{\bibfnamefont{A.}~\bibnamefont{Martin-Rodero}},
  \bibinfo{author}{\bibfnamefont{A.}~\bibnamefont{Levy~Yeyati}},
  \bibinfo{author}{\bibfnamefont{F.}~\bibnamefont{Flores}}, \bibnamefont{and}
  \bibinfo{author}{\bibfnamefont{R.~C.} \bibnamefont{Monreal}},
  \bibinfo{journal}{Phys. Rev. B} \textbf{\bibinfo{volume}{78}},
  \bibinfo{pages}{235112} (\bibinfo{year}{2008}),
  \urlprefix\url{http://link.aps.org/doi/10.1103/PhysRevB.78.235112}\relax
\mciteBstWouldAddEndPuncttrue
\mciteSetBstMidEndSepPunct{\mcitedefaultmidpunct}
{\mcitedefaultendpunct}{\mcitedefaultseppunct}\relax
\EndOfBibitem
\bibitem[{\citenamefont{Maier et~al.}(2011)\citenamefont{Maier, Schmidt, and
  Komnik}}]{PhysRevB.83.085401}
\bibinfo{author}{\bibfnamefont{S.}~\bibnamefont{Maier}},
  \bibinfo{author}{\bibfnamefont{T.~L.} \bibnamefont{Schmidt}},
  \bibnamefont{and} \bibinfo{author}{\bibfnamefont{A.}~\bibnamefont{Komnik}},
  \bibinfo{journal}{Phys. Rev. B} \textbf{\bibinfo{volume}{83}},
  \bibinfo{pages}{085401} (\bibinfo{year}{2011})\relax
\mciteBstWouldAddEndPuncttrue
\mciteSetBstMidEndSepPunct{\mcitedefaultmidpunct}
{\mcitedefaultendpunct}{\mcitedefaultseppunct}\relax
\EndOfBibitem
\bibitem[{\citenamefont{Galperin et~al.}(2007)\citenamefont{Galperin, Ratner,
  and Nitzan}}]{Galperin07}
\bibinfo{author}{\bibfnamefont{M.}~\bibnamefont{Galperin}},
  \bibinfo{author}{\bibfnamefont{M.~A.} \bibnamefont{Ratner}},
  \bibnamefont{and} \bibinfo{author}{\bibfnamefont{A.}~\bibnamefont{Nitzan}},
  \bibinfo{journal}{J. Phys.: Condens. Matter} \textbf{\bibinfo{volume}{19}},
  \bibinfo{pages}{103201} (\bibinfo{year}{2007})\relax
\mciteBstWouldAddEndPuncttrue
\mciteSetBstMidEndSepPunct{\mcitedefaultmidpunct}
{\mcitedefaultendpunct}{\mcitedefaultseppunct}\relax
\EndOfBibitem
\bibitem[{\citenamefont{Schmidt and Komnik}(2009)}]{PhysRevB.80.041307}
\bibinfo{author}{\bibfnamefont{T.~L.} \bibnamefont{Schmidt}} \bibnamefont{and}
  \bibinfo{author}{\bibfnamefont{A.}~\bibnamefont{Komnik}},
  \bibinfo{journal}{Phys. Rev. B} \textbf{\bibinfo{volume}{80}},
  \bibinfo{pages}{041307} (\bibinfo{year}{2009})\relax
\mciteBstWouldAddEndPuncttrue
\mciteSetBstMidEndSepPunct{\mcitedefaultmidpunct}
{\mcitedefaultendpunct}{\mcitedefaultseppunct}\relax
\EndOfBibitem
\bibitem[{\citenamefont{Haupt et~al.}(2009)\citenamefont{Haupt, Novotn\'y, and
  Belzig}}]{PhysRevLett.103.136601}
\bibinfo{author}{\bibfnamefont{F.}~\bibnamefont{Haupt}},
  \bibinfo{author}{\bibfnamefont{T.}~\bibnamefont{Novotn\'y}},
  \bibnamefont{and} \bibinfo{author}{\bibfnamefont{W.}~\bibnamefont{Belzig}},
  \bibinfo{journal}{Phys. Rev. Lett.} \textbf{\bibinfo{volume}{103}},
  \bibinfo{pages}{136601} (\bibinfo{year}{2009})\relax
\mciteBstWouldAddEndPuncttrue
\mciteSetBstMidEndSepPunct{\mcitedefaultmidpunct}
{\mcitedefaultendpunct}{\mcitedefaultseppunct}\relax
\EndOfBibitem
\bibitem[{\citenamefont{Avriller}(2011)}]{0953-8984-23-10-105301}
\bibinfo{author}{\bibfnamefont{R.}~\bibnamefont{Avriller}},
  \bibinfo{journal}{J. Phys. Condens. Matt.} \textbf{\bibinfo{volume}{23}},
  \bibinfo{pages}{105301} (\bibinfo{year}{2011})\relax
\mciteBstWouldAddEndPuncttrue
\mciteSetBstMidEndSepPunct{\mcitedefaultmidpunct}
{\mcitedefaultendpunct}{\mcitedefaultseppunct}\relax
\EndOfBibitem
\bibitem[{\citenamefont{Zazunov and Martin}(2007)}]{PhysRevB.76.033417}
\bibinfo{author}{\bibfnamefont{A.}~\bibnamefont{Zazunov}} \bibnamefont{and}
  \bibinfo{author}{\bibfnamefont{T.}~\bibnamefont{Martin}},
  \bibinfo{journal}{Phys. Rev. B} \textbf{\bibinfo{volume}{76}},
  \bibinfo{pages}{033417} (\bibinfo{year}{2007}),
  \urlprefix\url{http://link.aps.org/doi/10.1103/PhysRevB.76.033417}\relax
\mciteBstWouldAddEndPuncttrue
\mciteSetBstMidEndSepPunct{\mcitedefaultmidpunct}
{\mcitedefaultendpunct}{\mcitedefaultseppunct}\relax
\EndOfBibitem
\bibitem[{\citenamefont{Werner et~al.}(2009)\citenamefont{Werner, Oka, and
  Millis}}]{Werner2009}
\bibinfo{author}{\bibfnamefont{P.}~\bibnamefont{Werner}},
  \bibinfo{author}{\bibfnamefont{T.}~\bibnamefont{Oka}}, \bibnamefont{and}
  \bibinfo{author}{\bibfnamefont{A.~J.} \bibnamefont{Millis}},
  \bibinfo{journal}{Phys. Rev. B} \textbf{\bibinfo{volume}{79}},
  \bibinfo{pages}{035320} (\bibinfo{year}{2009})\relax
\mciteBstWouldAddEndPuncttrue
\mciteSetBstMidEndSepPunct{\mcitedefaultmidpunct}
{\mcitedefaultendpunct}{\mcitedefaultseppunct}\relax
\EndOfBibitem
\bibitem[{\citenamefont{Gull et~al.}(2011)\citenamefont{Gull, Millis,
  Lichtenstein, Rubtsov, Troyer, and Werner}}]{Werner2011}
\bibinfo{author}{\bibfnamefont{E.}~\bibnamefont{Gull}},
  \bibinfo{author}{\bibfnamefont{A.~J.} \bibnamefont{Millis}},
  \bibinfo{author}{\bibfnamefont{A.~I.} \bibnamefont{Lichtenstein}},
  \bibinfo{author}{\bibfnamefont{A.~N.} \bibnamefont{Rubtsov}},
  \bibinfo{author}{\bibfnamefont{M.}~\bibnamefont{Troyer}}, \bibnamefont{and}
  \bibinfo{author}{\bibfnamefont{P.}~\bibnamefont{Werner}},
  \bibinfo{journal}{Rev. Mod. Phys.} \textbf{\bibinfo{volume}{83}},
  \bibinfo{pages}{349} (\bibinfo{year}{2011})\relax
\mciteBstWouldAddEndPuncttrue
\mciteSetBstMidEndSepPunct{\mcitedefaultmidpunct}
{\mcitedefaultendpunct}{\mcitedefaultseppunct}\relax
\EndOfBibitem
\bibitem[{\citenamefont{M\"uhlbacher et~al.}(2011)\citenamefont{M\"uhlbacher,
  Urban, and Komnik}}]{PhysRevB.83.075107}
\bibinfo{author}{\bibfnamefont{L.}~\bibnamefont{M\"uhlbacher}},
  \bibinfo{author}{\bibfnamefont{D.~F.} \bibnamefont{Urban}}, \bibnamefont{and}
  \bibinfo{author}{\bibfnamefont{A.}~\bibnamefont{Komnik}},
  \bibinfo{journal}{Phys. Rev. B} \textbf{\bibinfo{volume}{83}},
  \bibinfo{pages}{075107} (\bibinfo{year}{2011})\relax
\mciteBstWouldAddEndPuncttrue
\mciteSetBstMidEndSepPunct{\mcitedefaultmidpunct}
{\mcitedefaultendpunct}{\mcitedefaultseppunct}\relax
\EndOfBibitem
\bibitem[{\citenamefont{Arrachea and Rozenberg}(2005)}]{PhysRevB.72.041301}
\bibinfo{author}{\bibfnamefont{L.}~\bibnamefont{Arrachea}} \bibnamefont{and}
  \bibinfo{author}{\bibfnamefont{M.~J.} \bibnamefont{Rozenberg}},
  \bibinfo{journal}{Phys. Rev. B} \textbf{\bibinfo{volume}{72}},
  \bibinfo{pages}{041301} (\bibinfo{year}{2005}),
  \urlprefix\url{http://link.aps.org/doi/10.1103/PhysRevB.72.041301}\relax
\mciteBstWouldAddEndPuncttrue
\mciteSetBstMidEndSepPunct{\mcitedefaultmidpunct}
{\mcitedefaultendpunct}{\mcitedefaultseppunct}\relax
\EndOfBibitem
\bibitem[{\citenamefont{Wang and Thoss}(2009)}]{Wang09}
\bibinfo{author}{\bibfnamefont{H.}~\bibnamefont{Wang}} \bibnamefont{and}
  \bibinfo{author}{\bibfnamefont{M.}~\bibnamefont{Thoss}},
  \bibinfo{journal}{The Journal of Chemical Physics}
  \textbf{\bibinfo{volume}{131}}, \bibinfo{eid}{024114}
  (pages~\bibinfo{numpages}{14}) (\bibinfo{year}{2009}),
  \urlprefix\url{http://link.aip.org/link/?JCP/131/024114/1}\relax
\mciteBstWouldAddEndPuncttrue
\mciteSetBstMidEndSepPunct{\mcitedefaultmidpunct}
{\mcitedefaultendpunct}{\mcitedefaultseppunct}\relax
\EndOfBibitem
\bibitem[{\citenamefont{Wang et~al.}(2011)\citenamefont{Wang, Pshenichnyuk,
  {H\"artle}, and Thoss}}]{Wang11}
\bibinfo{author}{\bibfnamefont{H.}~\bibnamefont{Wang}},
  \bibinfo{author}{\bibfnamefont{I.}~\bibnamefont{Pshenichnyuk}},
  \bibinfo{author}{\bibfnamefont{R.}~\bibnamefont{{H\"artle}}},
  \bibnamefont{and} \bibinfo{author}{\bibfnamefont{M.}~\bibnamefont{Thoss}},
  \bibinfo{journal}{J. Chem. Phys.} \textbf{\bibinfo{volume}{135}},
  \bibinfo{pages}{244506} (\bibinfo{year}{2011})\relax
\mciteBstWouldAddEndPuncttrue
\mciteSetBstMidEndSepPunct{\mcitedefaultmidpunct}
{\mcitedefaultendpunct}{\mcitedefaultseppunct}\relax
\EndOfBibitem
\bibitem[{\citenamefont{H\"{u}tzen et~al.}(2012)\citenamefont{H\"{u}tzen,
  Weiss, Thorwart, and Egger}}]{PhysRevB.85.121408}
\bibinfo{author}{\bibfnamefont{R.}~\bibnamefont{H\"{u}tzen}},
  \bibinfo{author}{\bibfnamefont{S.}~\bibnamefont{Weiss}},
  \bibinfo{author}{\bibfnamefont{M.}~\bibnamefont{Thorwart}}, \bibnamefont{and}
  \bibinfo{author}{\bibfnamefont{R.}~\bibnamefont{Egger}},
  \bibinfo{journal}{Phys. Rev. B} \textbf{\bibinfo{volume}{85}},
  \bibinfo{pages}{121408} (\bibinfo{year}{2012}),
  \urlprefix\url{http://link.aps.org/doi/10.1103/PhysRevB.85.121408}\relax
\mciteBstWouldAddEndPuncttrue
\mciteSetBstMidEndSepPunct{\mcitedefaultmidpunct}
{\mcitedefaultendpunct}{\mcitedefaultseppunct}\relax
\EndOfBibitem
\bibitem[{\citenamefont{Riwar and Schmidt}(2009)}]{Riwar2009}
\bibinfo{author}{\bibfnamefont{R.-P.} \bibnamefont{Riwar}} \bibnamefont{and}
  \bibinfo{author}{\bibfnamefont{T.~L.} \bibnamefont{Schmidt}},
  \bibinfo{journal}{Phys. Rev. B} \textbf{\bibinfo{volume}{80}},
  \bibinfo{pages}{125109} (\bibinfo{year}{2009})\relax
\mciteBstWouldAddEndPuncttrue
\mciteSetBstMidEndSepPunct{\mcitedefaultmidpunct}
{\mcitedefaultendpunct}{\mcitedefaultseppunct}\relax
\EndOfBibitem
\bibitem[{\citenamefont{Albrecht et~al.}(2012)\citenamefont{Albrecht, Wang,
  M\"uhlbacher, Thoss, and Komnik}}]{Ferdinand2012}
\bibinfo{author}{\bibfnamefont{K.~F.} \bibnamefont{Albrecht}},
  \bibinfo{author}{\bibfnamefont{H.}~\bibnamefont{Wang}},
  \bibinfo{author}{\bibfnamefont{L.}~\bibnamefont{M\"uhlbacher}},
  \bibinfo{author}{\bibfnamefont{M.}~\bibnamefont{Thoss}}, \bibnamefont{and}
  \bibinfo{author}{\bibfnamefont{A.}~\bibnamefont{Komnik}},
  \bibinfo{journal}{Phys. Rev. B} \textbf{\bibinfo{volume}{86}},
  \bibinfo{pages}{081412} (\bibinfo{year}{2012}),
  \urlprefix\url{http://link.aps.org/doi/10.1103/PhysRevB.86.081412}\relax
\mciteBstWouldAddEndPuncttrue
\mciteSetBstMidEndSepPunct{\mcitedefaultmidpunct}
{\mcitedefaultendpunct}{\mcitedefaultseppunct}\relax
\EndOfBibitem
\bibitem[{\citenamefont{{Gogolin} and {Komnik}}(2002)}]{Gogolin2002}
\bibinfo{author}{\bibfnamefont{A.~O.} \bibnamefont{{Gogolin}}}
  \bibnamefont{and} \bibinfo{author}{\bibfnamefont{A.}~\bibnamefont{{Komnik}}},
  \bibinfo{journal}{eprint arXiv:cond-mat/0207513}  (\bibinfo{year}{2002}),
  \eprint{arXiv:cond-mat/0207513}\relax
\mciteBstWouldAddEndPuncttrue
\mciteSetBstMidEndSepPunct{\mcitedefaultmidpunct}
{\mcitedefaultendpunct}{\mcitedefaultseppunct}\relax
\EndOfBibitem
\bibitem[{\citenamefont{Alexandrov et~al.}(2003)\citenamefont{Alexandrov,
  Bratkovsky, and Williams}}]{Alexandrov2003}
\bibinfo{author}{\bibfnamefont{A.~S.} \bibnamefont{Alexandrov}},
  \bibinfo{author}{\bibfnamefont{A.~M.} \bibnamefont{Bratkovsky}},
  \bibnamefont{and} \bibinfo{author}{\bibfnamefont{R.~S.}
  \bibnamefont{Williams}}, \bibinfo{journal}{Phys. Rev. B}
  \textbf{\bibinfo{volume}{67}}, \bibinfo{pages}{075301}
  (\bibinfo{year}{2003})\relax
\mciteBstWouldAddEndPuncttrue
\mciteSetBstMidEndSepPunct{\mcitedefaultmidpunct}
{\mcitedefaultendpunct}{\mcitedefaultseppunct}\relax
\EndOfBibitem
\bibitem[{\citenamefont{Galperin et~al.}(2005)\citenamefont{Galperin, Ratner,
  and Nitzan}}]{Galperin2005}
\bibinfo{author}{\bibfnamefont{M.}~\bibnamefont{Galperin}},
  \bibinfo{author}{\bibfnamefont{M.~A.} \bibnamefont{Ratner}},
  \bibnamefont{and} \bibinfo{author}{\bibfnamefont{A.}~\bibnamefont{Nitzan}},
  \bibinfo{journal}{Nano Letters} \textbf{\bibinfo{volume}{5}},
  \bibinfo{pages}{125} (\bibinfo{year}{2005})\relax
\mciteBstWouldAddEndPuncttrue
\mciteSetBstMidEndSepPunct{\mcitedefaultmidpunct}
{\mcitedefaultendpunct}{\mcitedefaultseppunct}\relax
\EndOfBibitem
\bibitem[{\citenamefont{D'Amico et~al.}(2008)\citenamefont{D'Amico, Ryndyk,
  Cuniberti, and Richter}}]{DAmico08}
\bibinfo{author}{\bibfnamefont{P.}~\bibnamefont{D'Amico}},
  \bibinfo{author}{\bibfnamefont{D.~A.} \bibnamefont{Ryndyk}},
  \bibinfo{author}{\bibfnamefont{G.}~\bibnamefont{Cuniberti}},
  \bibnamefont{and} \bibinfo{author}{\bibfnamefont{K.}~\bibnamefont{Richter}},
  \bibinfo{journal}{New Journal of Physics} \textbf{\bibinfo{volume}{10}},
  \bibinfo{pages}{085002} (\bibinfo{year}{2008}),
  \urlprefix\url{http://stacks.iop.org/1367-2630/10/i=8/a=085002}\relax
\mciteBstWouldAddEndPuncttrue
\mciteSetBstMidEndSepPunct{\mcitedefaultmidpunct}
{\mcitedefaultendpunct}{\mcitedefaultseppunct}\relax
\EndOfBibitem
\bibitem[{\citenamefont{Glazman and Shekhter}(1988)}]{Glazman1988}
\bibinfo{author}{\bibfnamefont{L.}~\bibnamefont{Glazman}} \bibnamefont{and}
  \bibinfo{author}{\bibfnamefont{R.}~\bibnamefont{Shekhter}},
  \bibinfo{journal}{Sov. Phys. JETP} \textbf{\bibinfo{volume}{67}},
  \bibinfo{pages}{163} (\bibinfo{year}{1988})\relax
\mciteBstWouldAddEndPuncttrue
\mciteSetBstMidEndSepPunct{\mcitedefaultmidpunct}
{\mcitedefaultendpunct}{\mcitedefaultseppunct}\relax
\EndOfBibitem
\bibitem[{\citenamefont{Mahan}(1991)}]{Mahan1991}
\bibinfo{author}{\bibfnamefont{G.}~\bibnamefont{Mahan}},
  \emph{\bibinfo{title}{Many-particle physics}} (\bibinfo{publisher}{Plenum
  press}, \bibinfo{year}{1991})\relax
\mciteBstWouldAddEndPuncttrue
\mciteSetBstMidEndSepPunct{\mcitedefaultmidpunct}
{\mcitedefaultendpunct}{\mcitedefaultseppunct}\relax
\EndOfBibitem
\bibitem[{\citenamefont{Schmidt et~al.}(2008)\citenamefont{Schmidt, Werner,
  M\"{u}hlbacher, and Komnik}}]{Schmidt2008}
\bibinfo{author}{\bibfnamefont{T.~L.} \bibnamefont{Schmidt}},
  \bibinfo{author}{\bibfnamefont{P.}~\bibnamefont{Werner}},
  \bibinfo{author}{\bibfnamefont{L.}~\bibnamefont{M\"{u}hlbacher}},
  \bibnamefont{and} \bibinfo{author}{\bibfnamefont{A.}~\bibnamefont{Komnik}},
  \bibinfo{journal}{Phys. Rev. B} \textbf{\bibinfo{volume}{78}},
  \bibinfo{eid}{235110} (\bibinfo{year}{2008})\relax
\mciteBstWouldAddEndPuncttrue
\mciteSetBstMidEndSepPunct{\mcitedefaultmidpunct}
{\mcitedefaultendpunct}{\mcitedefaultseppunct}\relax
\EndOfBibitem
\bibitem[{\citenamefont{Hewson and Newns}(1979)}]{hewson-newns}
\bibinfo{author}{\bibfnamefont{A.~C.} \bibnamefont{Hewson}} \bibnamefont{and}
  \bibinfo{author}{\bibfnamefont{D.~M.} \bibnamefont{Newns}},
  \bibinfo{journal}{Journal of Physics C: Solid State Physics}
  \textbf{\bibinfo{volume}{12}}, \bibinfo{pages}{1665} (\bibinfo{year}{1979}),
  \urlprefix\url{http://stacks.iop.org/0022-3719/12/i=9/a=009}\relax
\mciteBstWouldAddEndPuncttrue
\mciteSetBstMidEndSepPunct{\mcitedefaultmidpunct}
{\mcitedefaultendpunct}{\mcitedefaultseppunct}\relax
\EndOfBibitem
\bibitem[{\citenamefont{Lang and Firsov}(1963)}]{lang_firsov1963}
\bibinfo{author}{\bibfnamefont{I.~G.} \bibnamefont{Lang}} \bibnamefont{and}
  \bibinfo{author}{\bibfnamefont{Y.~A.} \bibnamefont{Firsov}},
  \bibinfo{journal}{Sov. Phys. JETP} \textbf{\bibinfo{volume}{16}},
  \bibinfo{pages}{1301} (\bibinfo{year}{1963})\relax
\mciteBstWouldAddEndPuncttrue
\mciteSetBstMidEndSepPunct{\mcitedefaultmidpunct}
{\mcitedefaultendpunct}{\mcitedefaultseppunct}\relax
\EndOfBibitem
\bibitem[{\citenamefont{Keldysh}(1964)}]{keldysh}
\bibinfo{author}{\bibfnamefont{L.}~\bibnamefont{Keldysh}},
  \bibinfo{journal}{Zh. Exp. Teor. Fiz.} \textbf{\bibinfo{volume}{47}}
  (\bibinfo{year}{1964}), \bibinfo{note}{[Sov. Phys. JETP {\bf 20}, 1018
  (1965)]}\relax
\mciteBstWouldAddEndPuncttrue
\mciteSetBstMidEndSepPunct{\mcitedefaultmidpunct}
{\mcitedefaultendpunct}{\mcitedefaultseppunct}\relax
\EndOfBibitem
\end{mcitethebibliography}

\end{document}